\documentclass[11pt]{article}
\usepackage{jheppub}

\usepackage[space]{grffile}
\usepackage{pdfpages}

\usepackage{slashed}
\usepackage{amsmath,amssymb,graphicx} 
\usepackage{epsf,color}
\usepackage{hyperref}
\usepackage{cancel}
\usepackage{framed}
\usepackage{hyperref}
\usepackage{float}
\usepackage{multirow}
\definecolor{nicered}{rgb}{0.5,0.,0.}
\definecolor{nicegreen}{rgb}{0.,0.5,0.}
\definecolor{niceblue}{rgb}{0.,0.,0.5}
\hypersetup{colorlinks,citecolor=nicegreen,linkcolor=nicered,urlcolor=niceblue}
\numberwithin{equation}{section}
\newcommand{\beq}{\begin{equation}}
	\newcommand{\eeq}{\end{equation}}
\newcommand{\bea}{\begin{eqnarray}}
	\newcommand{\eea}{\end{eqnarray}}
\newcommand{\bear}{\begin{eqnarray}}
	\newcommand{\eear}{\end{eqnarray}}

\newcommand{\GeV}{\textrm{GeV}}

\newcommand{\ba}{\begin{array}}
	\newcommand{\ea}{\end{array}}

\title{Heavy Neutral Leptons at the Electron-Ion Collider}

\author[a]{Brian Batell,}
\affiliation[a]{Pittsburgh Particle Physics, Astrophysics, and Cosmology Center, \\
	Department of Physics and Astronomy, University of Pittsburgh, Pittsburgh, USA}
\author[b]{Tathagata Ghosh,}
\affiliation[b]{Regional Centre for Accelerator-based Particle Physics, Harish-Chandra Research Institute,\\
	A CI of Homi Bhabha National Institute, Chhatnag Road, Jhusi, Prayagraj 211019, India}
\author[a]{Tao Han,}
\author[a]{and Keping Xie}

\emailAdd{batell@pitt.edu}
\emailAdd{than@pitt.edu}
\emailAdd{tathagataghosh@hri.res.in}
\emailAdd{xiekeping@pitt.edu}
\preprint{HRI-RECAPP-2022-006, PITT-PACC-2107}

\abstract{
	The future Electron-Ion Collider (EIC) at Brookhaven National Laboratory, along with its primary capacity to elucidate the nuclear structure, will offer new opportunities to probe physics beyond the Standard Model coupled to the electroweak sector. 
	Among the best motivated examples of such new physics are new heavy neutral leptons (HNLs), which are likely to play a key role in neutrino mass generation and lepton number violation. 
	We study the capability of the EIC to search for HNLs, which can be produced in electron-proton collisions 
	through charged current interactions as a consequence of their mixing with light neutrinos. We find that, with the EIC design energy and integrated luminosity, one is able to probe HNLs in the mass range of 
	1 GeV$-100$ GeV with mixing angles down to the order of $10^{-4} - 10^{-3}$ through the prompt decay, and $10^{-6} - 10^{-4}$ via the displaced decay signatures. We also consider the invisible mode where an HNL is undetected or decaying to dark sector particles. One could potentially probe heavy HNLs for mixing angles in the window $10^{-3}-10^{-2}$, provided SM background systematics can be brought under control. 
	These searches are complementary to other probes of HNLs, such as neutrino-less double-$\beta$ decay, meson decay, fixed-target, and high-energy collider experiments.
}

\begin{document}
	\maketitle
	
	\section{Introduction}
	\label{sec:intro}
	
	The Standard Model (SM) of elementary particle physics, 
	based on a non-Abelian SU(3)$_{\rm C}\otimes$ SU(2)$_{\rm L}\otimes$U(1)$_{\rm Y}$ gauge theory, 
	has been experimentally verified with a high 
	precision up to TeV-scale energies~\cite{Zyla:2020zbs}.  
	On the other hand, there is mounting evidence indicating the need for new physics beyond the SM 
	from disparate observations related to dark matter, neutrino mass generation, and matter/antimatter asymmetry, among others. Even within our physical realm at low energies, the luminous universe is predominantly made of 
	nucleons. Although it is understood that 
	the properties of the nucleons and nuclei are dictated by their quark and gluon constituents 
	and the SU(3)$_{\rm C}$ strong interaction of quantum chromodynamics (QCD) at low energies, there are still outstanding puzzles to be solved.
	The future Electron-Ion Collider (EIC)~\cite{Accardi:2012qut,AbdulKhalek:2021gbh}, to be built at  Brookhaven National Laboratory, will provide an unprecedented tool to explore the fundamental nature of nucleons and nuclei.
	The primary goals of the EIC physics program include the precise 3D tomographic imaging of partonic substructure, the determination of quark and gluon contributions to the proton spin, and the exploration of novel phases of nuclear matter at high densities. 
	To achieve these ends, the EIC will collide polarized electrons with polarized protons and ions over a wide range of energies and with high luminosities. Furthermore, access to a broad range of the partonic momentum fraction and momentum transfer $(x, Q^2)$ in the scattering processes will require a multipurpose hermetic detector with excellent tracking resolution and particle identification capabilities over a wide momentum region.
	
	The EIC will not only lead us to a new QCD frontier but will also have great potential to study precision electroweak (EW) physics and to search for new physics phenomena associated with the EW sector. These exciting prospects are a consequence of 
	the designed high luminosity, 
	relatively clean experimental environment in $eA$ collisions, and the multi-purpose detector design~\cite{AbdulKhalek:2021gbh,EIC-detector-handbook}.  
	Indeed, there are unique processes beyond the SM for EIC to explore~\cite{Kumar:2016mfi}.
	First, the precision determinations for the EW neutral current~\cite{Boughezal:2022pmb} and the weak mixing angle~\cite{Kumar:2013yoa,Kumar:2016mfi,Zhao:2016rfu} will  
	provide sensitive probes of new light neutral gauge boson interactions ($Z'$)~\cite{RamseyMusolf:1999qk,Davoudiasl:2012ag,Erler:2014fqa,Yan:2022npz}. 
	The intense incoming electron beam provides a good laboratory for searching for charged lepton-flavor transition. A unique signature will be a leptoquark state~\cite{Gonderinger:2010yn}, 
	or analogously an $R$-parity violating interaction ($\lambda'$) in Supersymmetry (SUSY), most readily 
	produced in the $s$-channel in lepton-quark collisions if kinematically accessible~\cite{Adloff:1999tp,Chekanov:2005au}. 
	Recent studies have also highlighted the promising sensitivity of the EIC to axion-like particles (ALPs)~\cite{Davoudiasl:2021mjy,Liu:2021lan}.
	Much more work and new ideas are needed to expand the new physics coverage potentially accessible at the EIC. 
	
	In this paper, we explore another class of new physics signatures from a new heavy neutral lepton (HNL), denoted $N$~\cite{Abdullahi:2022jlv}.  HNLs are a common feature of many extensions 
	of the SM, motivated by their role in addressing the generation of neutrino masses.  The best-known model including $N$ is the Type-I Seesaw mechanism for neutrino mass~\cite{Minkowski:1977sc,Yanagida:1979as,GellMann:1980vs,Glashow:1979nm,Mohapatra:1979ia,Schechter:1980gr}, and its variations~\cite{Mohapatra:1986aw,Mohapatra:1986bd,Bernabeu:1987gr,Malinsky:2005bi}. In the standard Type-I scenario, there exists a Majorana mass term and neutrinos are thus all Majorana. The smoking-gun signature would be a lepton-number violation by two units. The neutrino-less double-beta decay experiments have been the dedicated driver in the search lepton number violation for decades~\cite{Kim:2020vjv}. Meson decays~\cite{Atre:2009rg} and collider searches for lepton-number violation are being actively carried out~\cite{Cai:2017mow}. In some other scenarios, the heavy neutrino may be (quasi-)Dirac without the observable effect of lepton-number violation~\cite{Valle:1983dk}. 
	From a phenomenological point of view, we choose the HNL mass and the mixing elements to be free parameters without specifying any underlying model.
	We set out to identify the experimental signatures, quantify the signal and backgrounds, and estimate the achievable sensitivities to HNLs at the EIC. 
	Our search strategies are generally applicable to other new physics searches at the EIC involving final states of charged leptons and jets, both prompt and displaced, and may provide some general guidance for future considerations.

	The rest of the paper is organized as follows. 
	We first present a brief overview of the EIC to set the stage in Sec.~\ref{sec:EIC}, including the relevant collider parameters and detector capabilities. We then introduce the HNL model along with a description of the relevant production and decay processes 
	in Sec.~\ref{sec:N} to guide our studies.   In Sec.~\ref{sec:Simul}, we describe in detail our simulation methodology and HNL search strategies at the EIC, for a variety of signals governed by the mixing and mass parameters and the corresponding backgrounds.  For completeness, we also list the current search and bounds on the model parameters.
	We summarize our results and offer further discussions and an outlook in Sec.~\ref{sec:Disc}.

	\section{The Electron-Ion Collider}
	\label{sec:EIC} 
	
	The Electron-Ion Collider (EIC) is designed to study the properties of the nucleons and nuclei with unprecedented precision. 
	As argued in the introduction, the powerful beam and detector capabilities of the EIC also afford exciting opportunities to
	probe a variety of new physics beyond the SM.
	
	The EIC will utilize the existing Relativistic Heavy Ion Collider (RHIC) 
	facility with its two intersecting superconducting rings, each 3.8 km in circumference. 
	A polarized electron beam with an energy up to 21 GeV will be set to collide with a number of ion species accelerated in the existing RHIC accelerator complex, from polarized protons with a peak energy of 250 GeV to fully stripped uranium ions with energies up to 100 GeV/u, covering a center-of-mass (c.m.)~energy range from 30 to 145 GeV for polarized $ep$, and from 20 to 90 GeV for $eA$ 
	(for a large $A$)~\cite{Accardi:2012qut}. 
	The maximum beam energy could be further increased by about $10\%$. 
	Using one of the two RHIC hadron rings and the Energy Recovery Linac (ERL) as the electron accelerator, the EIC could reach a high luminosity in the $10^{33} - 10^{34}$ cm$^{-2}$s$^{-1}$ range.
	For our analyses in this work, we choose the following benchmark for the c.m.~energy and  integrated luminosity for $ep$ collisions as
	\beq
	\sqrt s = 141\ {\rm GeV,}\quad \mathcal{L}=100\ {\rm fb}^{-1}.
	\label{eq:sqrt-s}
	\eeq
	Polarizations of $70\%$ may be achievable for the electron and nucleon beams, and this will be relevant when we consider the production of HNLs below. 
	
	To achieve the rich physics program of the EIC, a high-performance multi-purpose detector is required to accommodate the extended interaction region for a wide range in c.m.~energy, different combinations of beam particle species, and a broad variety of distinct physics processes. 
	The various physics processes encompass inclusive and semi-inclusive measurements induced via neutral current and charged current interactions,
	\beq
	e + p/A \to e' + X,\quad
	e + p/A \to \nu_e + X,
	\eeq
	where $X$ generically denotes any observable leptons/hadrons as well as the beam remnants. 
	The detector requirements include a good tracking system, electromagnetic and hadronic calorimetry, a muon chamber, good hermetic coverage, as well as vertex determination. For further details on the EIC detector capabilities, see Ref.~\cite{AbdulKhalek:2021gbh}.
	
	\section{Heavy Neutral Leptons} 
	\label{sec:N}
	
	Heavy neutral leptons (HNLs, $N$) are a common feature 
	in many extensions of the SM. They are particularly motivated by the need for new dynamics associated with neutrino masses, as in the Type-I Seesaw mechanism~\cite{Minkowski:1977sc,Yanagida:1979as,GellMann:1980vs,Glashow:1979nm,Mohapatra:1979ia,Schechter:1980gr}, and light HNLs near the weak scale may also play a role in the generation of the matter-antimatter asymmetry~\cite{Asaka:2005pn,Akhmedov:1998qx} or provide a portal to thermal relic dark matter~\cite{Pospelov:2007mp,Bertoni:2014mva,Gonzalez-Macias:2016vxy,Escudero:2016ksa,Batell:2017rol,Batell:2017cmf,Schmaltz:2017oov}.
	In the standard Type-I scenario, there exists a Majorana mass term, $M N^2$, and as a result, both the light neutrinos and HNLs are Majorana particles. 
	A smoking-gun signature of this scenario would be a lepton-number violation by two units. 
	However, in some other scenarios, the heavy neutrino may be (quasi-)Dirac without the observable effect of lepton-number violation~\cite{Mohapatra:1986aw,Mohapatra:1986bd,Bernabeu:1987gr,Malinsky:2005bi}. 
	As we will demonstrate, direct searches for HNLs can be carried out in both scenarios at the EIC.  We set out to identify the experimental signatures, quantify the signal and backgrounds, and estimate the achievable sensitivities at the EIC. 
	
	HNLs couple to SM through the neutrino portal operator, 
	\beq
	-\mathcal L\supset y_{\nu}^{iI} \,   {\hat L_i}  \, H   {\hat N_I}+{\rm{H.c.}}\,,
	\label{eq:LHN}
	\eeq
	where $H$ is the SM Higgs doublet and ${\hat L_i}=(\hat\nu_i,\hat\ell_i)^T$ is the SM leptonic flavor doublet with $i = e,\mu,\tau$. 
	The index $I$ in Eq.~(\ref{eq:LHN}) runs over the number of HNLs present in the theory. 
	Note that we use 2-component Weyl spinors here.
	Following electroweak symmetry breaking, the HNLs will mix with the SM neutrinos. In the mass basis, the HNL interactions with the SM particles are governed by the mixing matrix, $U$, and are given as 
	\begin{equation}
		\label{eq:HNL-weak-interaction}
		{\cal L} \supset \frac{g}{\sqrt{2}} \, U_{iI}  \, W_\mu^- \, \ell_i^\dag  \, \overline \sigma^\mu N_I  + \frac{g}{2 \, c_W} U_{iI} \,   Z_\mu \,   \nu_i^\dag \,\overline \sigma^\mu N_I  +{\rm H.c.}
	\end{equation}
	In our setting, we choose the HNL mass $m_N$ and the mixing elements $U_{iI}$ to be free phenomenological parameters without specifying any underlying model.
	The production of HNLs and their subsequent decays thus depend on their induced couplings to electroweak bosons with strengths controlled by $U_{iI}$. 
	As such, the leading production mechanism for $N$ at the EIC will proceed via the charged current interaction
	\beq
	e + p/A \to N + X.
	\label{eq:eN}
	\eeq
	In particular, the production is governed by the strength of electron-flavor mixing $U_{eI}$. 
	With our primary aim of characterizing the EIC prospects for probing HNLs, we follow a simplified approach and assume that a single HNL $(I = 1)$ with electron-flavor mixing dominance, \emph{i.e.}, $U_e \neq 0$ while $U_\mu = U_\tau =0$, is present in the $1 - 100$ GeV mass range.
	The parameter space is then completely characterized by the HNL mass, $m_N$, and the mixing angle, $U_e$. 
	We will explore both cases of Majorana and Dirac HNLs in our study. 
	
	\begin{figure}[tb]
		\begin{center}
			\includegraphics[width=.51\textwidth]{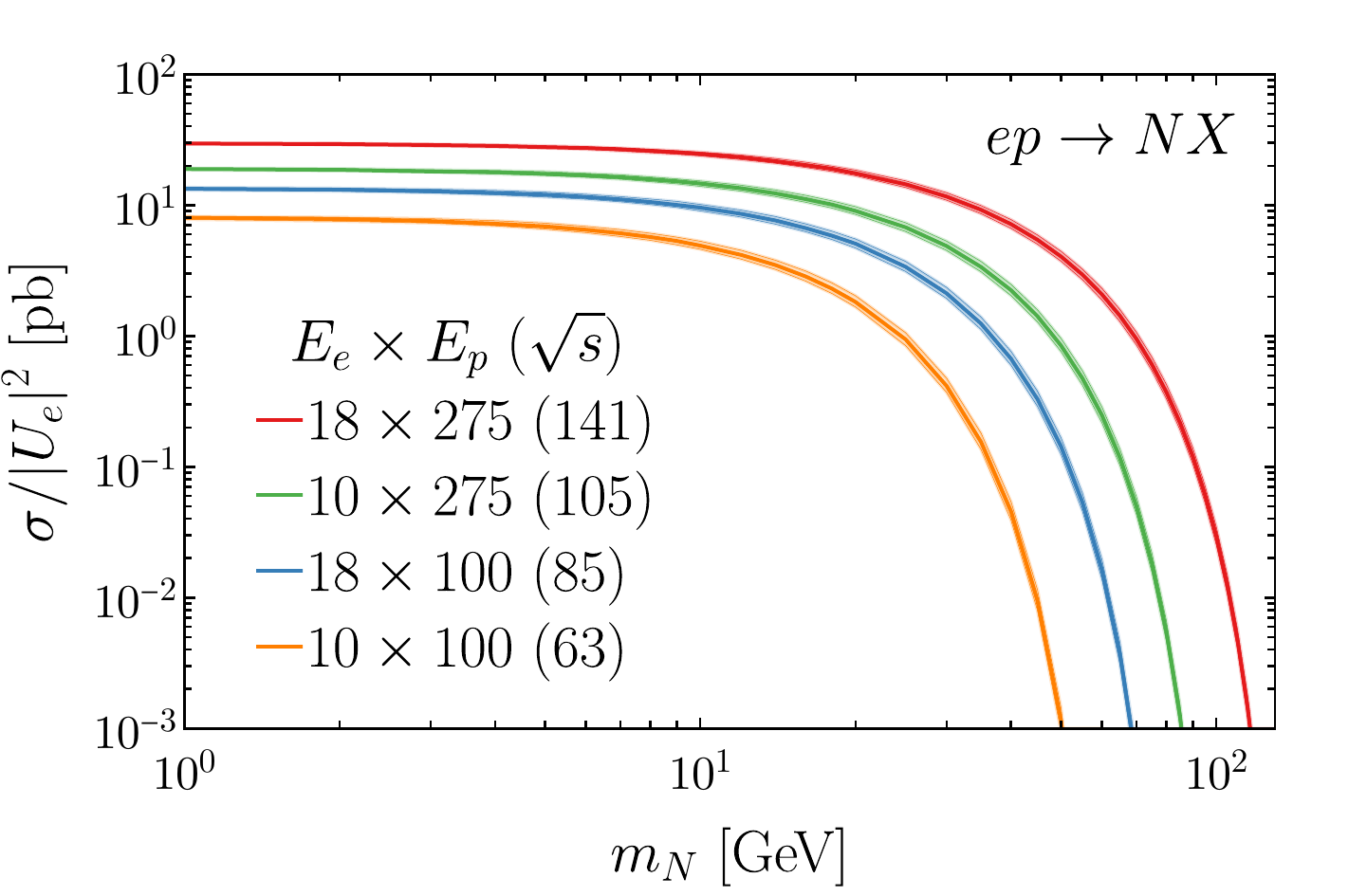}
			\includegraphics[width=.48\textwidth]{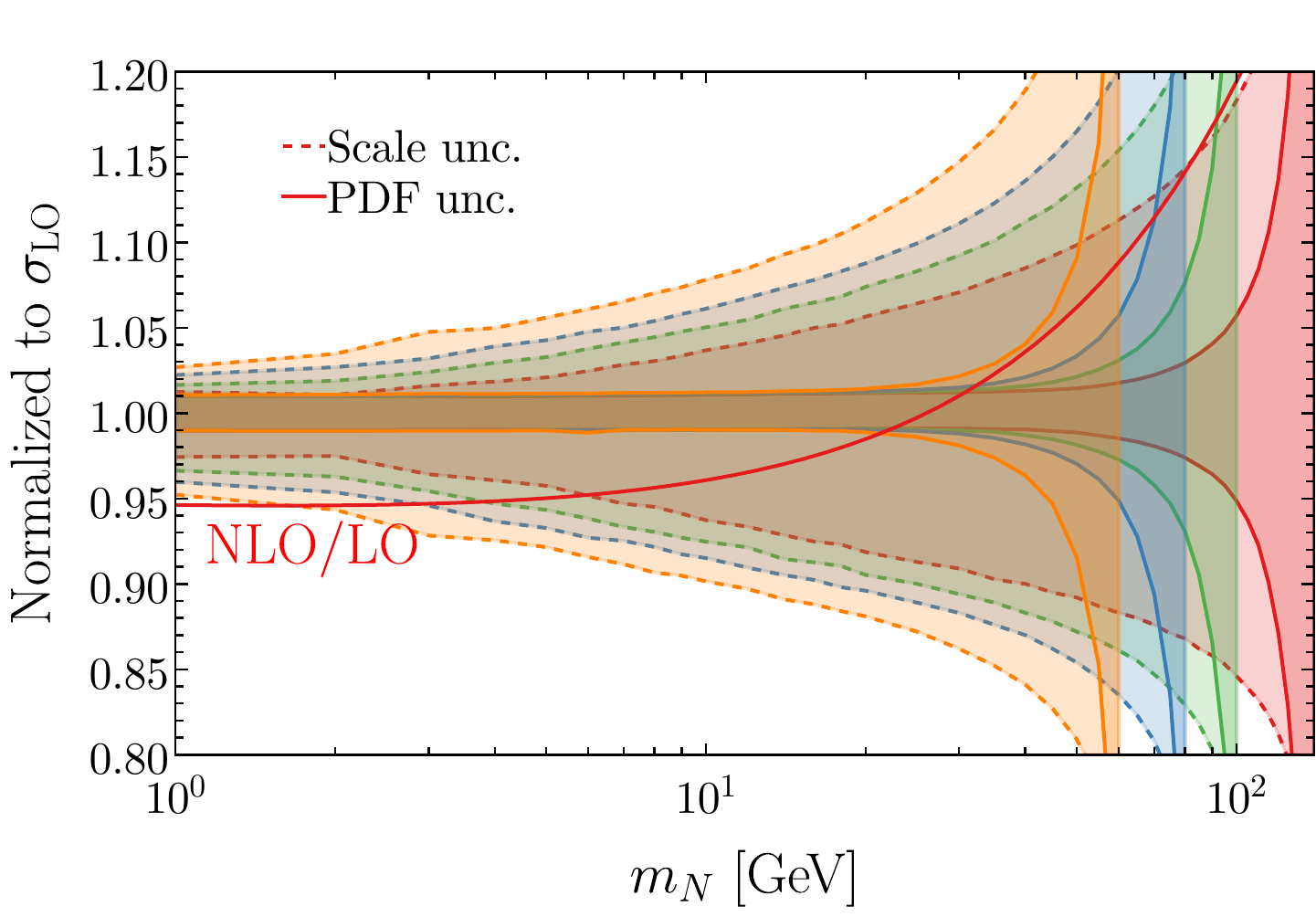}
		\end{center}
		\caption{Left: The HNL production cross sections divided by the squared mixing parameter $|U_e|^2$ at the EIC with unpolarized beam energies as $E_p\times E_e~[\GeV^2]~(\sqrt{s}=\sqrt{4E_e E_p}~[\GeV])$ versus the HNL mass. Right: The scale and PDF error bands for the HNL cross section at the EIC, estimated with CT18NNLO PDFs and varying renormalization and factorization scales by a factor of 2. 
			The orange solid line indicates the NLO/LO cross section ratio for the representative $\sqrt{s}=141~\GeV$ machine. 
		}
		\label{fig:xsec}
	\end{figure}
	
	In Fig.~\ref{fig:xsec}, we present the HNL production cross sections  
	including next-to-leading order (NLO) QCD corrections (factoring out the squared mixing parameter $|U_e|^2$) as a function of the HNL mass $m_N$ for $ep$ collisions with beam energies as $E_{p}=100,275$ GeV and $E_e=10,18$ GeV, respectively, adopting the CT18NNLO parton distribution functions \cite{Hou:2019efy}. 
	Similar calculations have been performed in $ep$ collisions at the HERA energies  \cite{Buchmuller:1990vh,Buchmuller:1991tu,Ingelman:1993ve} as well as for the proposed LHeC~\cite{Das:2018usr,Li:2018wut,Gu:2022muc} and beam dump experiments at future lepton colliders~\cite{Giffin:2022rei}.
	The production cross sections are the same for the Dirac and Majorana HNLs, as they share the same gauge couplings 
	in Eq.~(\ref{eq:HNL-weak-interaction}). 
	We see that for low masses the production cross sections are essentially constant in HNL mass for the assumed collider energies, while they decrease sharply for heavier masses near the threshold due to the kinematic suppression.
	The default renormalization and factorization scales are chosen as $\mu_{R,F}^{} =\sqrt{Q^2+m_N^2}$, where $Q$ is the momentum transfer of the incoming electron, $Q^2=-(p_e-p_N)^2$. The scale uncertainty is estimated by varying $\mu_{R,F}$ by a factor of 2. It is found to be a few percent when $m_N\lesssim10$ GeV, and 20\% when $m_N\sim100$ GeV, shown in Fig.~\ref{fig:xsec} (right).
	In comparison, the PDF uncertainty is typically at a few-percent level.
	Both scale and PDF relative uncertainties gradually increase with the HNL mass as a natural result of the decrease of the reference absolute cross sections.
	
	We also notice that NLO QCD high-order corrections to the cross sections\footnote{The NLO infrared safe cross section requires a well-defined jet. We take the anti-$k_T$ algorithm with $p_T^j>5~\GeV$ and $\Delta R=0.4$, which applies both to the LO and NLO when obtaining the ratio. The calculation is done with \texttt{Sherpa}~\cite{Sherpa:2019gpd}. } are about $-5\%$ in the GeV mass region while $+20\%$ around $m_N \sim 100$ GeV, shown as the red solid line in Fig.~\ref{fig:xsec} (right) for the $18\times275~\GeV^2$ collision, with the corresponding scale uncertainty expected to be reduced. 
	In our exploration of unknown new physics, these theoretical uncertainties are not expected to play a significant role in the sensitivity reach. 
	We will thus adopt the LO calculation in our following simulations, assuming that theoretical uncertainties are understood adequately. 
	
	One of the important features of the EIC is the electron beam polarization. This is particularly advantageous when probing new physics with chiral couplings. 
	Assuming the electron beam to have a percentage longitudinal polarization $P$ with $P=-1$ as purely left-handed and $P=+1$ as purely right-handed, we have the polarized cross section as 
	\beq
	\sigma(P) = {1\over 2}[ (1-P)\sigma_- +  (1+P)\sigma_+] .
	\eeq
	In the SM, the charged current interaction is left-handed, so that the total cross section reaches the maximum for the purely left-handed beam $\sigma(P=-1) = \sigma_-=2\sigma(P=0)$, while the cross section vanishes for the purely right-handed beam $\sigma(P=+1) = \sigma_+=0$. In the following, we follow the EIC Yellow Report~\cite{AbdulKhalek:2021gbh} and adopt $P=-70\%$ for the electron beam. As a result, the polarized cross section will be magnified by a factor of 1.7 compared with the unpolarized cross section shown in Fig.~\ref{fig:xsec}.
	We note that the EIC proton beam can be polarized up to 70\% as well. However, the parton polarization difference $\Delta f_i=f_i^{+}-f_i^{-}$ is generally small compared with its average  $f_i=(f_i^++f_i^-)/2$~\cite{Nocera:2014gqa}. Therefore, we use the unpolarized PDFs. Meanwhile, we will mainly focus on the proton beam throughout this work, which gives a larger cross section with respect to the ion beams, because of the larger collision energy as well as richer up-quark parton components.
	
\begin{figure}
		\centering
		\includegraphics[width=0.3\textwidth]{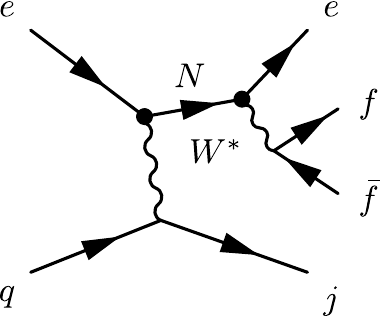}
		\includegraphics[width=0.3\textwidth]{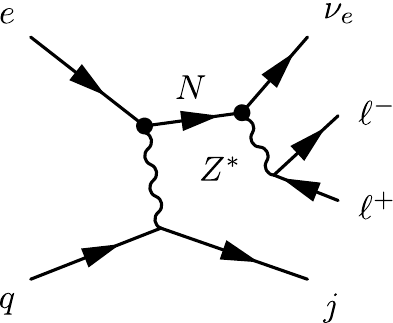}
		\caption{Feynman diagrams for $N$ production and decays 
			via charged (left) and neutral (middle) currents and a three-body far-off shell. 
		}
		\label{fig:feyn}
	\end{figure}

	Once produced, $N$ will subsequently decay via the charged and neutral current processes
	\beq
	N \to e W^{(*)},\quad \nu_e Z^{(*)},
	\eeq
	with the subsequent decays of the gauge bosons $W/Z$ to a pair of fermions. 
	When $m_N$ is below the $W/Z$ threshold, the HNL can decay to three-body final states mediated via virtual $W/Z$ bosons. 
	We depict the representative Feynman diagrams for the production and decay in
	Fig.~\ref{fig:feyn}.
	The total decay width for the three-body decay (Majorana type) can be estimated as
	\begin{equation}\label{eq:DW}
		\Gamma_N\sim\frac{G_F^2m_N^5}{192\pi^3}|U_e|^2\sum_{i=\ell,q} N_{c}^i\Theta(m_N-m_{X}^i)C_V^i.
	\end{equation}
	Using the results of Refs.~\cite{Gorbunov:2007ak,Canetti:2012kh,Shuve:2016muy,Bondarenko:2018ptm} to sum over all the channels $i$ gives an overall factor of $\sum_i\sim23.8$, which depends on the hadronization of quark final states.
	In our analysis, we neglect the hadronization effects but restore the threshold effect in each channel $i$, and sum all open ones with $\Theta(m_N-m_{X}^i)$, which is a unit (zero) when the channel is open (forbidden). The color factor $N_{c}^i$ takes 3 (1) for hadronic (leptonic) channels. 
	The coupling factor $C^i_V$ depends on the isospins and charges of the final-state particles, as well as the gauge boson mediators, $W^*$ or/and $Z^*$.
	
	\begin{figure}[tb]
		\begin{center}
			\includegraphics[width=0.53\textwidth]{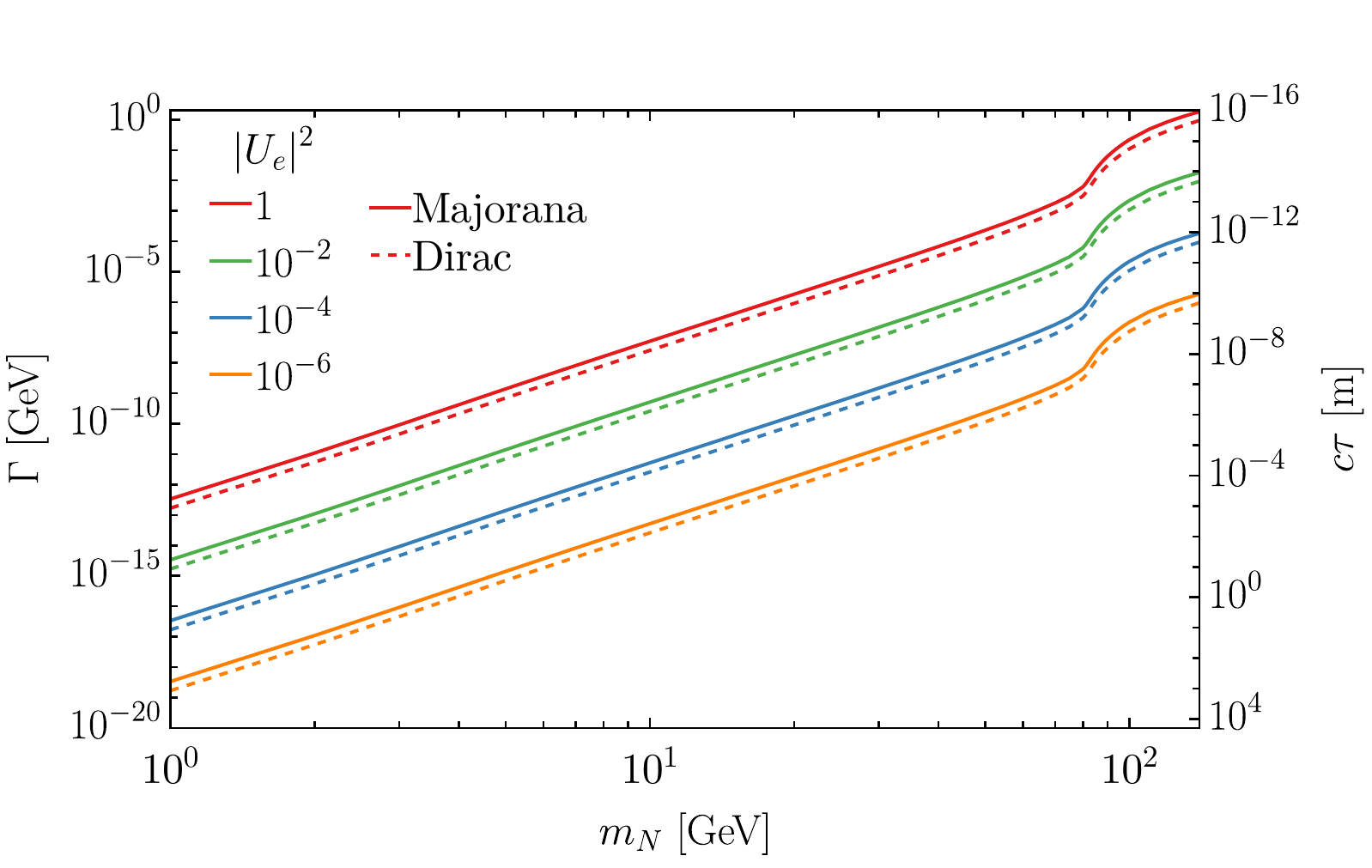}
			\includegraphics[width=0.46\textwidth]{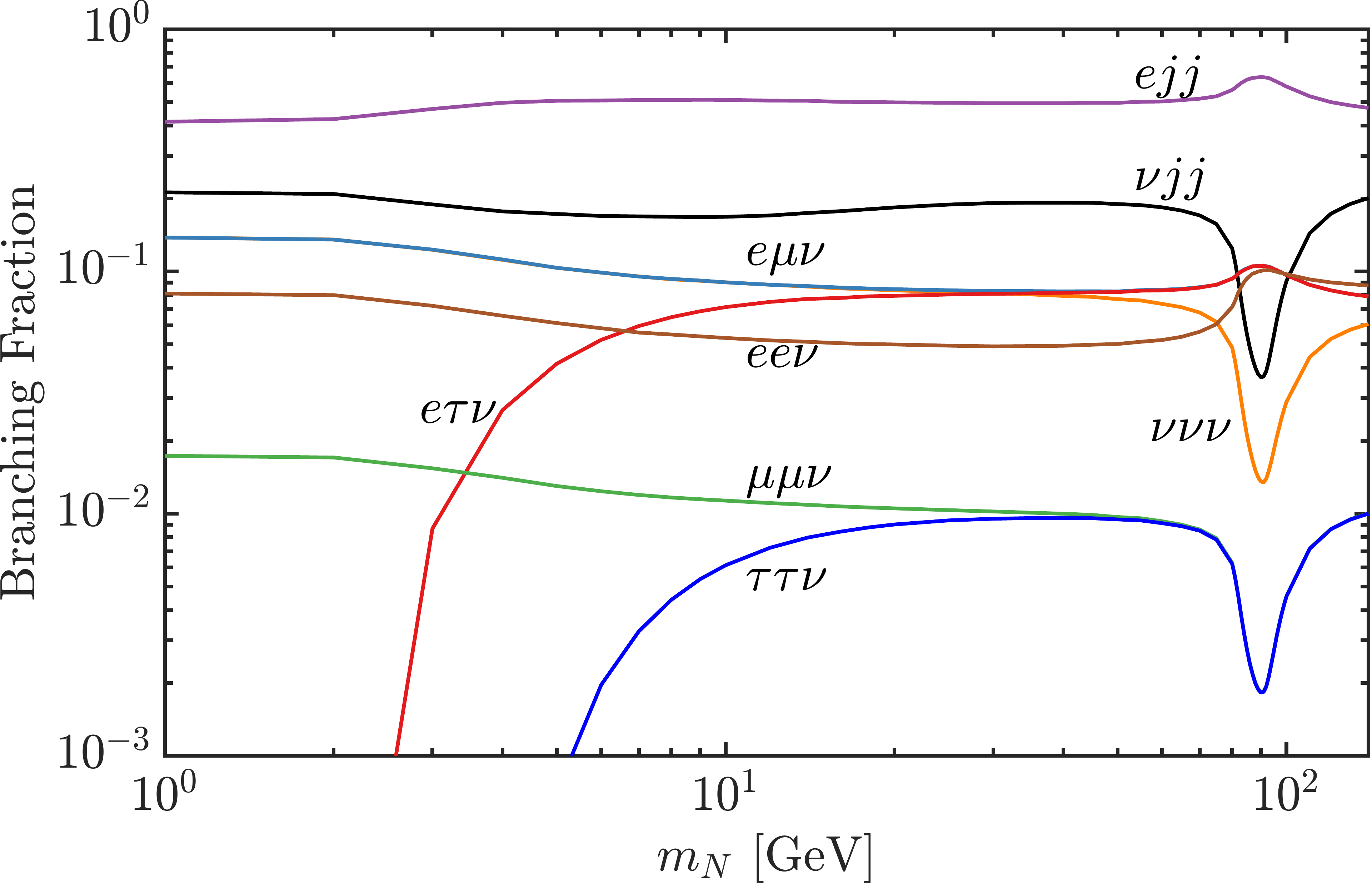}
		\end{center}
		\caption{The HNL total decay width (left) and the branching fractions to final states as labeled (right). 
		}
		\label{fig:BrN}
	\end{figure}
	
	We show the decay width versus $m_N$ for several values of $|U_e|^2$ in Fig.~\ref{fig:BrN} (left). 
	We see that the decay width of the Majorana HNL (solid curves) is twice that of the Dirac type (dashed curves) because the Majorana HNL can decay through both lepton number violating and conserving modes,  
	\emph{i.e.}, $N\to e^{\pm}W^{\mp},\overset{(-)}{\nu_e}Z$. The width has a sharp increase above the $W/Z$ threshold. 
	Using the decay width in Eq.~(\ref{eq:DW}) for $m_N \lesssim m_W$, we estimate the HNL proper lifetime as
	\begin{equation}\label{eq:tauN}
		\tau_N =\frac{1}{\Gamma_N}\sim\  10^{-9}\ {\rm s} \times \left({1\ {\rm GeV} \over m_N}\right)^5 \left( { 10^{-3} \over |U_e|^2} \right).
	\end{equation}
	We indicate the proper decay length $c\tau$ by the vertical axis on the right. We see that for small mixing $|U_e|^2$, the HNL decay width can be very small and $N$ can be long-lived.  For instance, considering $m_N<10$ GeV and $|U_e|^2<10^{-4}$, the HNL decay length could be of order 100 $\mu$m or larger.
	As a result, the experimental signatures can be quite different, as we will discuss in the following section. 
	In Fig.~\ref{fig:BrN} (right), we show the $N$ decay branching fractions to the fermionic states. We see that the channels from the $eW$ mode are 2$-$10 times as much as the corresponding ones from the $\nu Z$ mode, depending on the specific fermions in the final state and their respective thresholds.
	When $m_N$ crosses the $W$ threshold, the two-body decay channels open, and the decay width increase drastically.
	As a result, the branching fractions of the $eZ$ channels present sharp dips, before rising again after $m_N$ crosses the $Z$ threshold. 
	While we are particularly interested in the $e,\mu$ final states from the observational point of view, the $\tau$ final state may be also of interest above the $m_\tau$ threshold. 
	We note that in the di-lepton channels, the $ee\nu$ branching is smaller than the $e\mu\nu$ channel, as a result of the destructive interference between the $Z^{*}$ and $W^{*}$ mediated diagrams, as depicted in Fig.~\ref{fig:feyn}. When $m_N\gg m_\tau$, the $e\tau\nu$ shares roughly the same branching as the $e\mu\nu$, reflecting the lepton universality.
	
	\section{Simulation and Analysis}
	\label{sec:Simul}
	
	With an understanding of the production and decays of the HNLs in hand, we now describe the simulations and analyses that will be used to derive our sensitivity projections for HNL searches at the EIC. We will consider several classes of HNL signatures in this section.
	As shown in Fig.~\ref{fig:BrN} (right), for visible decays to SM final states there are multiple decay channels available to search for the $N$ signal. We choose to focus on the decay channels with $e$ and $\mu$ in the final states for a clear signal identification and background suppression. We also consider the situation in which $N$ is long-lived, leading to a displaced decay as a unique signal. We finally explore a more challenging scenario in which $N$ decays to invisible final-state particles, which may occur if $N$ has additional decay modes to invisible dark particles. 
	
	\begin{table}[!htp]
		\begin{center}
			\begin{tabular}{|c| c|}
				\hline
				$\eta$ & Resolution\\
				\hline     
				\hline
				\multicolumn{2}{|c|}{Tracking ($\sigma_{p}/p$)}\\
				\hline
				$ 2.5 < |\eta| \leq 3.5$ &  $ 0.1\% \times p \, \oplus \, 2\%$  \\
				$ 1.0 < |\eta| \leq 2.5$ &  $ 0.02\% \times p \, \oplus \, 1\%$  \\
				$  |\eta| \leq 1.0$ &  $ 0.02\% \times p \, \oplus \, 5\%$  \\
				\hline
				\hline
				\multicolumn{2}{|c|}{Electromagnetic calorimeter ($\sigma_{E}/E$)}\\
				\hline 
				$ -3.5 \leq \eta < -2.0$ &  $ 1\%/E \, \oplus 2.5\%/ \sqrt{E} \, \oplus \, 1\%$  \\
				$ -2.0 \leq \eta < -1.0$ & $ 2\%/E \, \oplus 6\%/ \sqrt{E} \, \oplus \, 2\%$  \\
				$ -1.0 \leq \eta < 1.0$ & $ 2\%/E \, \oplus 13\%/ \sqrt{E} \, \oplus \, 2.5\%$  \\
				$  1.0 \leq \eta \leq 3.5$ &   $ 2\%/E \, \oplus 8\%/ \sqrt{E} \, \oplus \, 2\%$ \\
				\hline 
				\multicolumn{2}{|c|}{Hadronic calorimeter ($\sigma_{E}/E$)}\\
				\hline 
				$ 1.0 < |\eta| \leq 3.5$ &  $ 50\%/ \sqrt{E}\, \oplus \, 10\%$  \\
				$ |\eta| \leq 1.0$ &  $ 100\%/ \sqrt{E}\, \oplus \, 10\%$  \\
				\hline    
			\end{tabular}
		\end{center}
		\caption{\label{tab:DetectorMatrix} Angular coverage, tracking momentum resolution, and calorimeter resolutions of the EIC detector used in our analysis. These parameters are based on~\cite{EIC-detector-handbook,Arratia:2020azl}.}
	\end{table}

	\subsection{Prompt HNL searches}
	We first consider searches for promptly decaying HNLs.\,\footnote{The prompt searches are simulated with the beam profile as $E_e\times E_p=20\times250~\GeV^2$~\cite{Accardi:2012qut}, slightly different from the design in the recent EIC Yellow Report~\cite{AbdulKhalek:2021gbh}, though with the same c.m. energy $\sqrt{s} = 141$ GeV as in Eq.~(\ref{eq:sqrt-s}). This will not significantly impact our sensitivity projections.} 
	For the sake of clean experimental observation at the EIC, we will only consider electrons and muons ($\ell=e,\mu$), unless explicitly stated. We discuss three distinct prompt leptonic decay channels of the HNL. The first two analyses target lepton number violating (LNV) channels to search for a Majorana HNL: $e^- p \to e^+ \, + \, 3j$ which is the classic LNV channel of HNL searches \cite{Buchmuller:1991tu,Duarte:2014zea,Antusch:2016ejd}, and $e^- p \to  e^+\mu^- j + \, E_T^{\text{miss}}$. The third analysis focuses on a Dirac HNL in the channel $e^- p \to\ell^-\ell^+j + E_T^{\text{miss}}$. 
	
	We take the UFO model files of Majorana and Dirac HNLs from Refs.~\cite{Alva:2014gxa,Degrande:2016aje,Pascoli:2018heg}. 
	The signal and background events are simulated by using \texttt{MadGraph5\_aMC} v2.6.7~\cite{Alwall:2014hca} with the CT18NNLO parton distribution functions~\cite{Hou:2019efy}. Thereafter, we pass our simulated events through a toy detector before analyzing them. We develop our toy detector code based on the angular coverages and resolutions of the tracker and electromagnetic and hadronic calorimeters from the EIC Yellow report, as summarized in Table~\ref{tab:DetectorMatrix}.

	\begin{figure}
		\centering
		\includegraphics[width=0.49\textwidth]{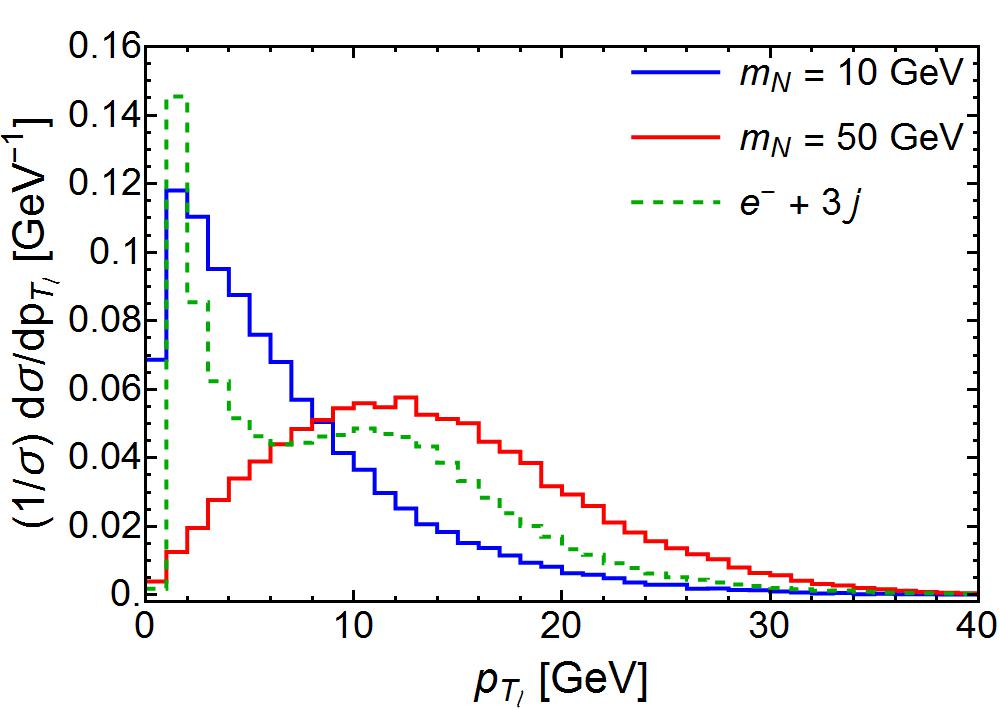}
		\includegraphics[width=0.48\textwidth]{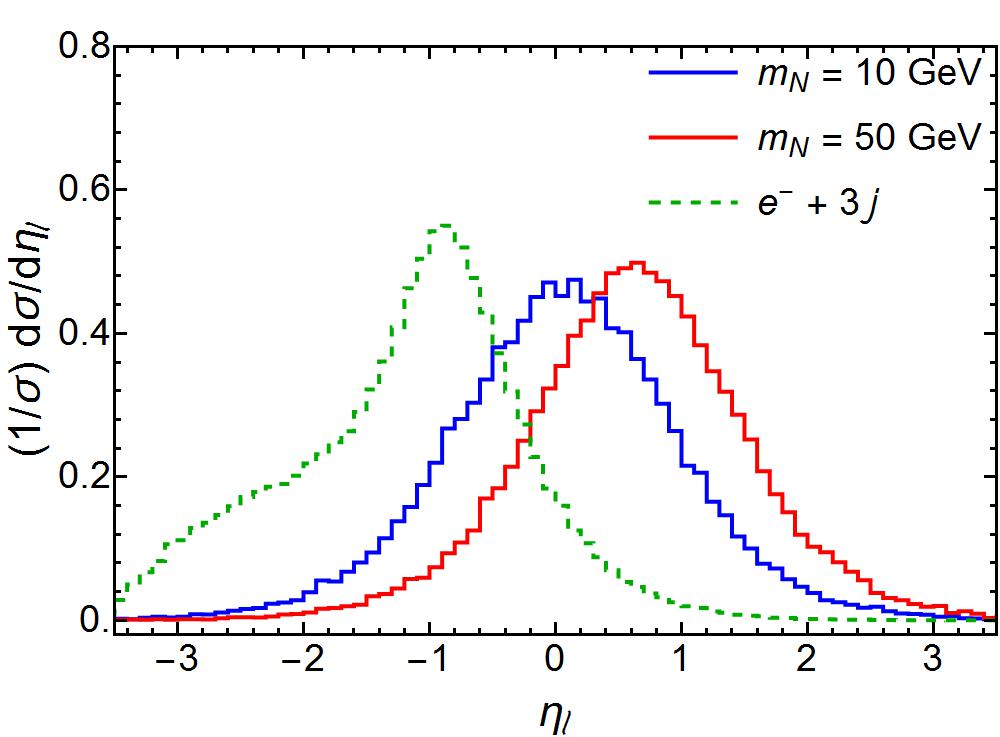}
		\includegraphics[width=0.49\textwidth]{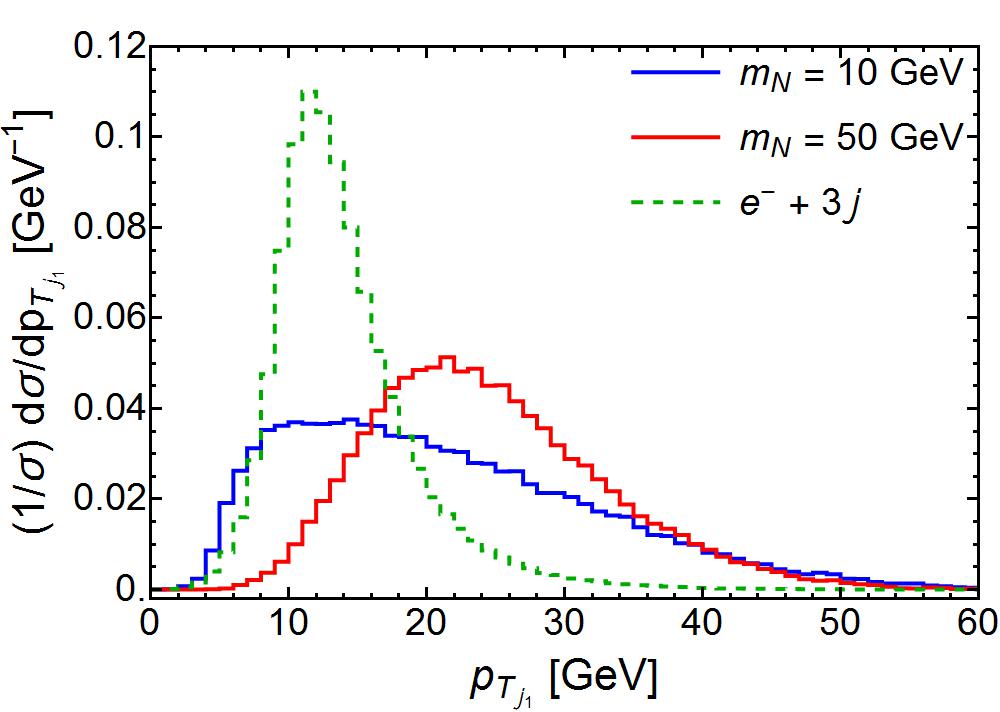}
		\includegraphics[width=0.48\textwidth]{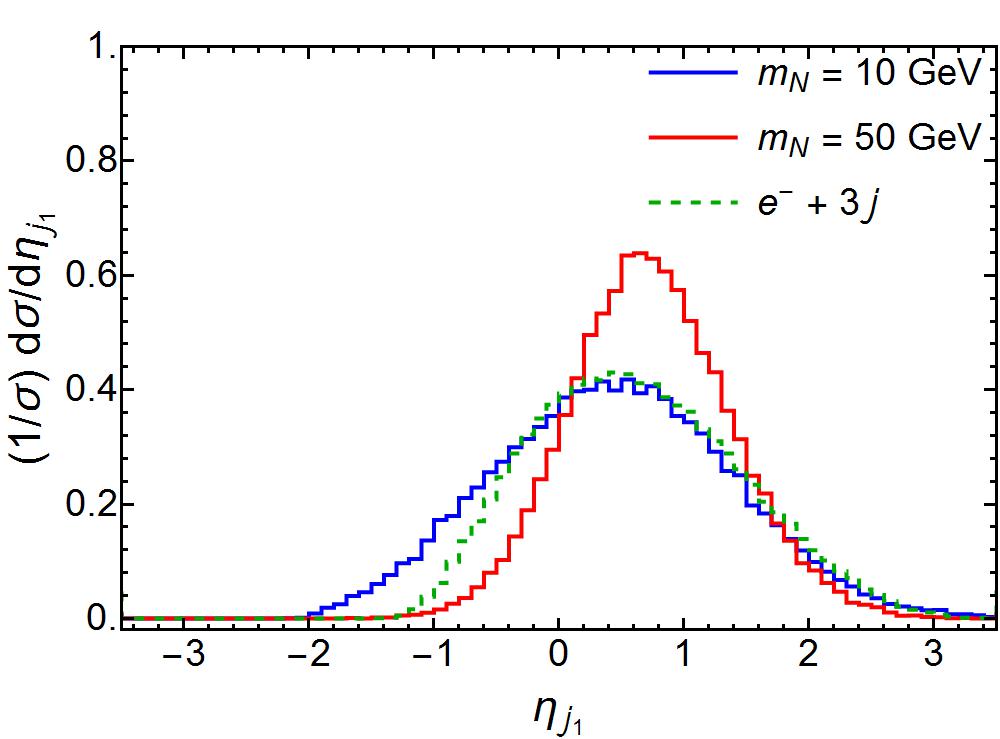}
		\caption{The transverse momentum (left panels)  and pseudo-rapidity (right panels) distributions of the lepton (top panels) and the leading jet (bottom panels) of the signal and the $e^-+3j$ background in the $e^-p\to  e^+jjj$ channel before applying the cuts of Eqs.~(\ref{eq:lepcut}) and (\ref{eq:jetcut}).  
		}
		\label{fig:dist_e+jjj_pre_lepton}
	\end{figure}

	\subsubsection{$e^++3j$ search for Majorana HNL}
	
	We first consider the semi-leptonic decay channel of the HNL $N\to e^+(W^{-(*)}\to2j)$, leading to the lepton-number-violating signal $e^+ + 3j$. This is a genuine $\Delta L=2$ process, and there is no irreducible background for $e^+$ production in the SM. The fake backgrounds include the pair production of $\gamma^*\to e^+e^-$ with $e^-$ missing from detection, and the neutral current process of $e^-+$jets with $e^-$ misidentified as an $e^+$. Although the fake rates would be low, such as (0.01--0.1)\%, the production rates still are high due to the large cross section of the neutral-current deep-inelastic scattering. Therefore, without pursuing more detailed optimization including the detection issues for those rather inclusive processes, and to be conservative, we will focus on the exclusive signal from $e^+ + 3j$.
	
	For this analysis, the cut-flow strategy for this final state is straightforward and is shown in Table~\ref{tab:cut-flow_Majorana_LNV} for two representative HNL masses, $m_N= 10$ GeV and 50 GeV with $|U_e|^2=1$. 
	For this study, we select exactly one isolated lepton with\footnote{For efficient simulation of signal and background events we use $p_{T_{\ell}} > 1$ GeV at the generator level. In contrast, for the jets, we impose $p_{T_{j}} > 5 \, (1)$ GeV for the signal (background) generation. For further optimization of the background sample production, a stronger $p_{T}$ cut of 10 GeV is used on only the leading jet. Finally, for background events, only isolated jets are simulated by using $\Delta R (\ell / j,j) > 0.4$. }
	
	\begin{equation}\label{eq:lepcut}
		p_{T_{\ell}} > 2\ {\rm GeV\ and}\  0 < \eta_{\ell} < 3.5. 
	\end{equation} 
	Also, we select three jets within 
	\begin{equation}\label{eq:jetcut}
		|\eta_j| < 3.5\ {\rm with}\ \ p_{T_{j_1}} > 20\ {\rm GeV,\ and}\ p_{T_{j_{2,3}}} > 5\ {\rm GeV}.
	\end{equation}
	The transverse momentum and pseudo-rapidity distributions of the lepton and the leading jet, before applying the above cuts are shown in Fig.~\ref{fig:dist_e+jjj_pre_lepton}. It is worthwhile to point out that in Eq.~(\ref{eq:lepcut}), we select leptons only with positive $\eta$ values. The preference of final state leptons to be in the forward hemisphere, especially those arising from heavier HNL decays, is quite evident from the top right panel of Fig.~\ref{fig:dist_e+jjj_pre_lepton}. Hence, together, the cuts of Eq.~(\ref{eq:lepcut}) reduce the background by an order of magnitude, with signal efficiencies being $40 - 80 \%$ depending on the HNL mass. In contrast, for the leading jet pseudo-rapidity, the signal from lower HNL masses is not clearly distinguishable from the background, and thus, we keep jets in both hemispheres to increase signal acceptances. However, one can see from the lower left panel of Fig.~\ref{fig:dist_e+jjj_pre_lepton} that the $p_T$ of the leading jet of the background peaks between $10-15$ GeV but the signal $p_{T_{j_1}}$ distributions are pretty broad.  
	Thus, the $p_{T_{j_1}} > 20$ GeV cut suppresses the background by another factor of 30 but reduces the signal by a factor of 2-3 only. To ensure that the lepton and jets are isolated, we impose 
	\begin{equation}\label{eq:drlj}
		\Delta R (\ell,j_{\alpha}) > 0.4 \ {\rm and}\ \Delta R (j_{\alpha},j_{\beta}) > 0.4 \, (\alpha,\beta=1,2,3).
	\end{equation} 
	The isolation requirements can be adjusted and optimized once the detector performance is better understood.
	
	\begin{table}[tb]
		\begin{center}
			\resizebox{\columnwidth}{!}{
				\begin{tabular}{|c|c c| c|}
					\hline
					\multirow{3}{*}{ Cut selection} & \multicolumn{2}{c|}{Signal $[e^- p\to(N\to e^+ jj)j]$} & \multirow{2}{*}{$e^- jjj$}\\
					\cline{2-3}
					& $m_N = 10$ GeV & $m_N = 50$ GeV &  \\ 
					& [pb] & [pb] & [pb] \\
					\hline   
					\hline
					Production & 5.53 & 0.95 & 449 \\
					\hline 
					Exactly $1 \ell$: &  \multirow{2}{*}{2.43} & \multirow{2}{*}{0.74} & \multirow{2}{*}{36.7}\\
					$p_{T_{\ell}} > 2$ GeV, $0 < \eta_{\ell} < 3.5$ & & &\\
					\hline
					Exactly $3 j$: &  \multirow{2}{*}{0.81} & \multirow{2}{*}{0.43} & \multirow{2}{*}{1.35}\\
					{$p_{T_{j_1}} > 20$ GeV}, $p_{T_{j_{2,3}}} > 5$ GeV, $|\eta_{j_{1,2,3}}| < 3.5$ & & & \\
					\hline
					Isolation: &  \multirow{2}{*}{0.22} & \multirow{2}{*}{0.39} & \multirow{2}{*}{1.35}\\
					$\Delta R (\ell/j_{\alpha},j_{\beta}) > 0.4~(\alpha,\beta=1,2,3)$ & & &\\
					\hline
					\multirow{2}{*}{$\Delta M^{\rm min} =\min\big(|M(\ell j_{\alpha}j_{\beta}) -m_N|\big) <5$ GeV}
					& 0.22 & $\times$ & 0.03\\
					& $\times$ & 0.30 & 0.64 \\
					\hline
					\hline
					\multirow{2}{*}{Require one $e^+$ [$f^{\text{MID}}=0.1 \%$]} & 0.22 &  $\times$ & $3.23 \times 10^{-5}$\\
					& $\times$ & 0.30 & $6.40 \times 10^{-4}$ \\
					\hline
					\multirow{2}{*}{Require one $e^+$ [$f^{\text{MID}}=0.01 \%$]} & 0.22 &  $\times$ & $3.23 \times 10^{-6}$\\  
					& $\times$ & 0.30 & $6.40 \times 10^{-5}$ \\
					\hline
					Polarization $P_e=-70\%$ & $\times1.7$ & $\times1.7$  & $\times1$ \\
					\hline
				\end{tabular}
			}
		\end{center}
		\caption{\label{tab:cut-flow_Majorana_LNV} Cut-flow table of the Majorana HNL signal, with $|U_e|^2=1$ in the $ e^+ + 3j $ final state. The last row indicates the cross-section enhancement factor for a $P_e=-70\%$ polarized electron beam. Similarly for the tables below.
		}
	\end{table}
	
\begin{figure}
		\centering
		\includegraphics[width=0.49\textwidth]{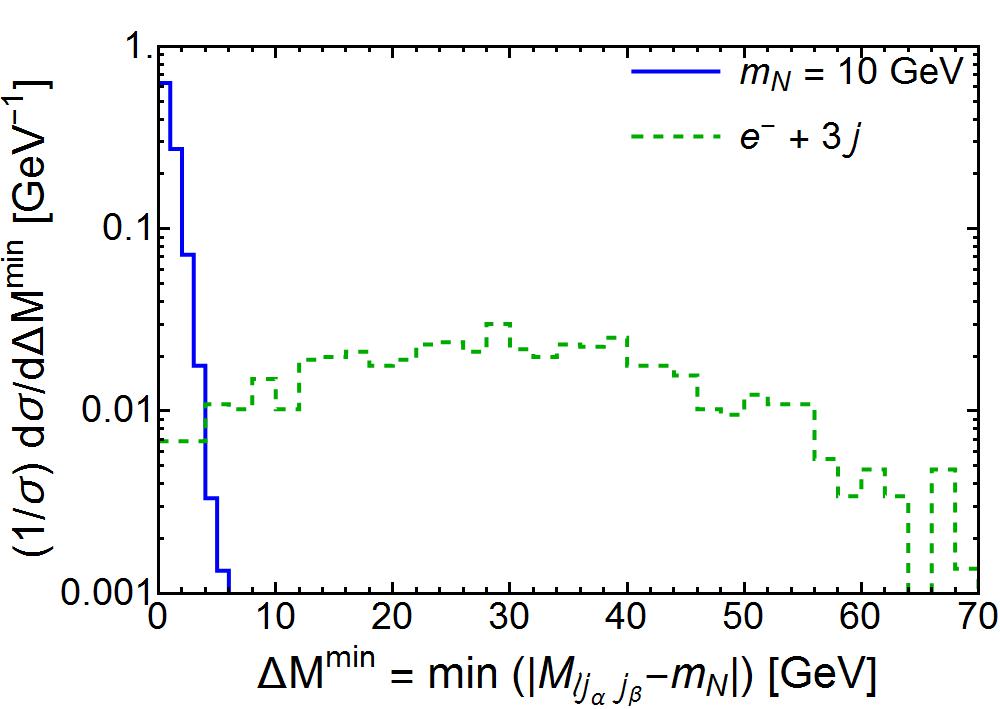}
		\includegraphics[width=0.49\textwidth]{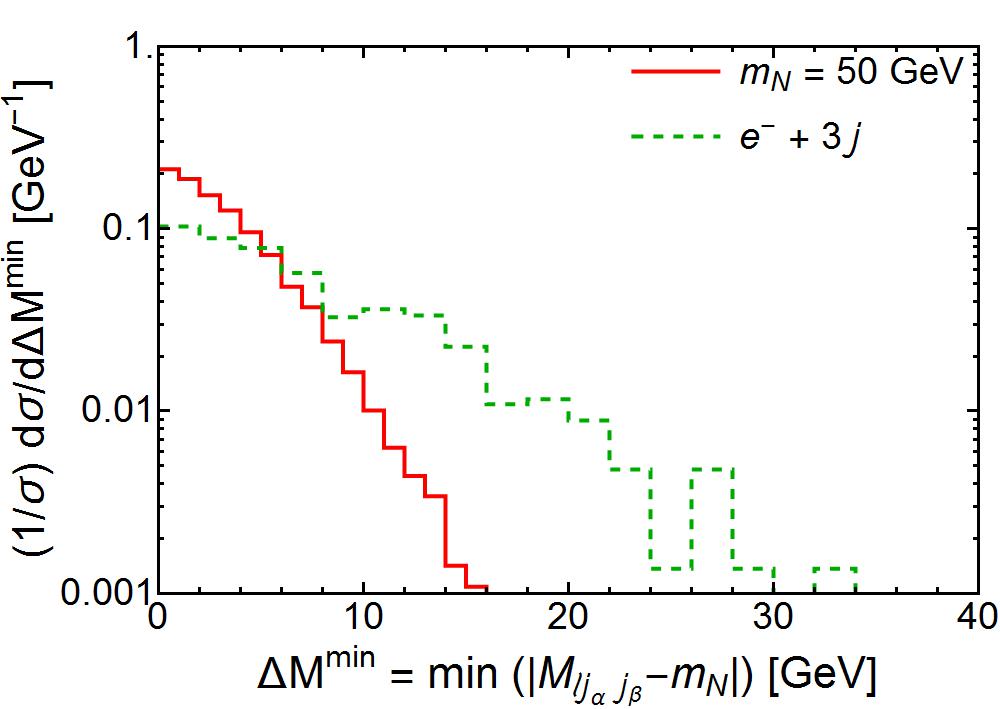}
		\caption{$\Delta M^{\text{min}}$ distributions for $m_N = 10$ (50) GeV along with the same for the $e^- +3j$ backgrounds, after applying the cuts of Eq.~(\ref{eq:drlj}).}
		\label{fig:dist_e+jjj_post_isolation}
	\end{figure}

	
	For the signal search with a hypothetical mass $m_N$,
	we construct the variable $\Delta M^{\text{min}} = |M(\ell j_{\alpha}j_{\beta}) -m_N|$ where $\alpha < \beta=1,2,3$, which give us three values, and we require the minimum of those three to be less than 5 GeV. It should be noted that this variable is a function of the HNL mass and affects the background differently for different HNL masses. For $m_N=10$ and 50 GeV, this cut suppresses the background by factors of 20 and 2, respectively. The corresponding signal efficiencies are $100\%$ and $78\%$. The $\Delta M^{\text{min}}$ distributions of signals with $m_N=10$ (left) and 50 GeV (right) against the corresponding backgrounds are shown in Fig.~\ref{fig:dist_e+jjj_post_isolation}.

	Ultimately, we require the selected lepton to be a positron, and apply the electron (positron) charge misidentification rate ($f^{\text{MID}}$) to the remaining $e^- + 3j$ background to estimate its contribution to our analysis. Motivated by the recent estimations from the LHC~\cite{ATLAS:2019qmc},\footnote{In Ref.~\cite{ATLAS:2019qmc} the ATLAS collaboration has shown that by using tight identification and isolation criteria for electrons, and utilising a BDT, the electron charge misidentification rate can be as low as $\lesssim 0.1 \, (0.2) \%$ for $E_{T_e} < 60 \, (80)$ GeV. However, it gradually rises to $\sim 1 \%$ for $E_{T_e} > 200$ GeV. In our present analysis, the electrons typically have $p_T <50$ GeV, as is evident from the top left panel of Fig.~\ref{fig:dist_e+jjj_pre_lepton}. Hence, we use $f^{\text{MID}} = 0.1 \%$ in our study.} first we use $f^{\text{MID}} = 0.1 \%$ to obtain $S/B \sim 10^4 \,\, (10^2)$ with $|U_e|^2=1$ for $m_N = 10 \, \, (50)$ GeV benchmark. 
	Naturally, we also note that the EIC will be cleaner than the gluon-rich environment of the LHC, and hence, one can expect that the charge of an electron can be determined more accurately at the EIC, leading to a reduced $f^{\text{MID}}$. With this motivation, we have also performed the analysis with a more optimistic assumption of $f^{\text{MID}}= 0.01 \%$, which leads to another order of magnitude improvement in $S/B$ for this analysis. 
	
	\subsubsection{$\mu^- e^+ j \, + \, E_T^{\text{miss}}$ search for Majorana HNL}
	\label{sec:prompt_emu_Majorana}
	
	We next focus on another LNV channel $e^- p \to e^+ \mu^- \bar\nu_\mu \, j$.
	The HNL signal arises via $N\to e^+(W^{-(*)}\to\mu^-\bar\nu_\mu)$. 
	This channel is essentially SM background free, if we are able to effectively identify the lepton-number violating decay $N\to e^+ W^{-(*)}$.
	Although observing an isolated $e^+$ is a good start for the signal identification, some fake backgrounds from $\gamma^*/Z^{*}\to e^+e^-$ and  $e^-+$\,jets may have a large production rate. Again, we will not pursue the optimal search for the rather inclusive $e^+$ signal. We focus on the dilepton channel $\mu^-e^+$ in order to exclude SM backgrounds originated from $\gamma^{*}/Z^{*}\to\ell^+\ell^-$, which give the same-flavor lepton pairs.
	
	For the exclusive signal under consideration,
	the only significant background for this analysis is from the cascade decay $\gamma p\to \tau^- \tau^+ j \rightarrow \mu^- e^+ j \, + \, 4 \nu$, where the photon is radiated by the incoming electron. 
	We follow the standard treatment of the equivalent photon approximation~\cite{Budnev:1975poe}. 
	The scattered beam electron will be lost along the beam pipe.
	
	For our analysis, we select exactly two charged leptons with the acceptance cuts 
	\begin{equation}\label{eq:lepcut2}
		p_{T_{\ell}} > 2\ {\rm GeV\ and}\   |\eta_{\ell}| < 3.5.
	\end{equation} 
	For the jet, we take
	\begin{equation}\label{eq:jetcut2}
		p_{T_{j}} > 10\ {\rm GeV\ and}\  |\eta_{j}| < 3.5, 
	\end{equation} 
	followed by the isolation criteria involving the leptons and the jet given below\footnote{For the simulations of both signal and background samples pertaining to this analysis, we employ $p_{T_{\ell (j)}} > 1 \, (5)$ GeV at the generator level.}
	\begin{equation}\label{eq:drlj2}
		\Delta R (\ell_1, \ell_2) > 0.3\ {\rm and}\ \Delta R (\ell_{1,2},j) > 0.4.
	\end{equation} 
	%
	
	\begin{figure}
		\centering
		\includegraphics[width=0.49\textwidth]{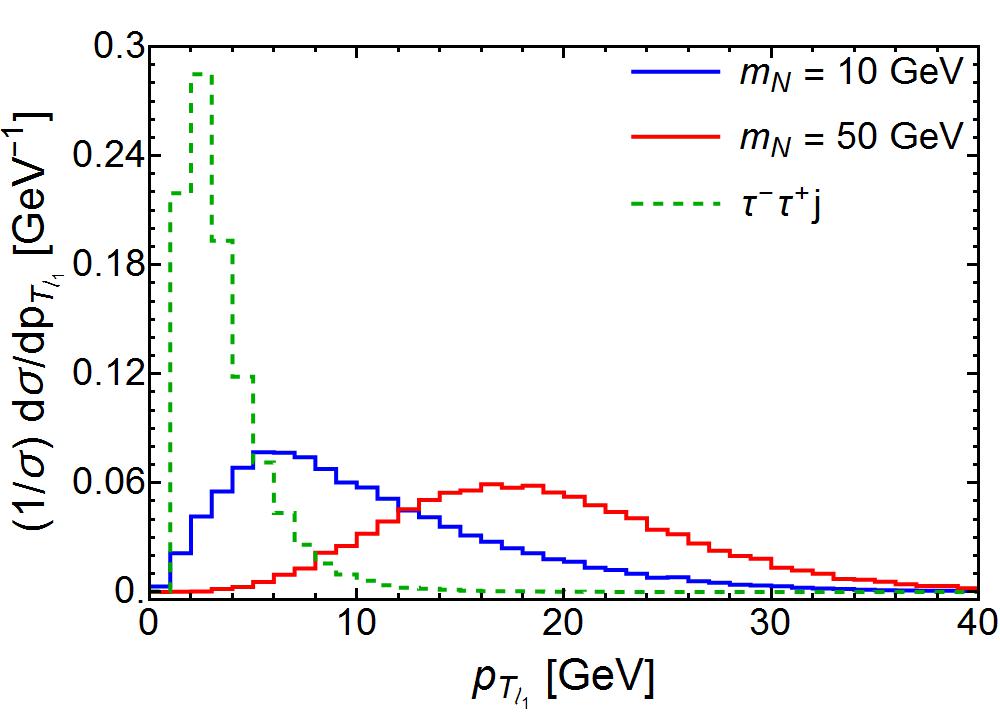}
		\includegraphics[width=0.49\textwidth]{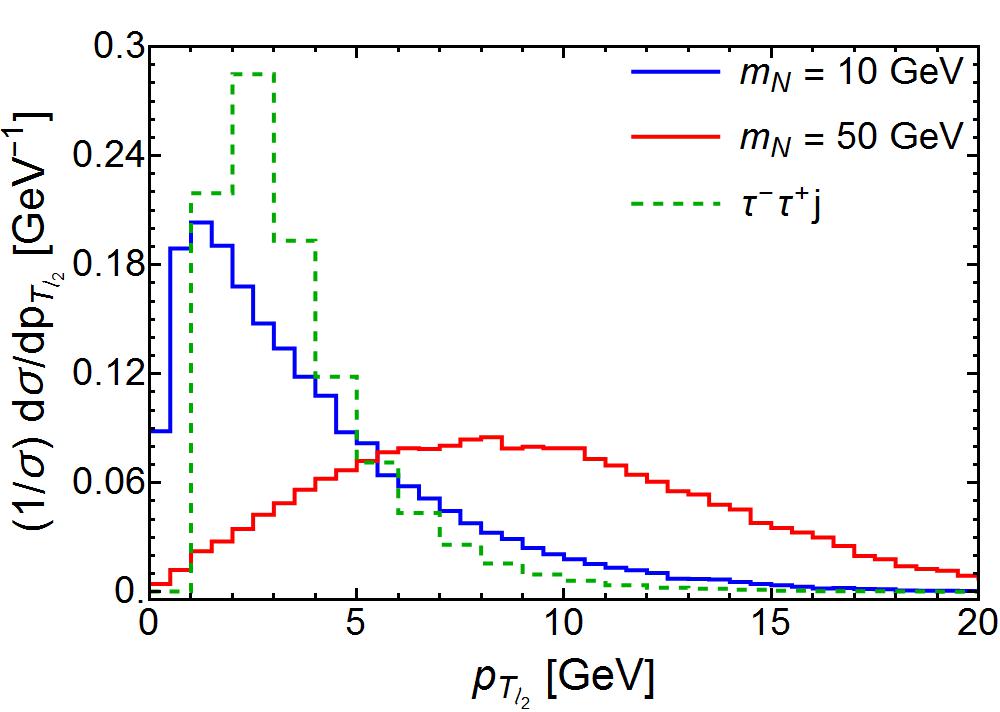}
		\includegraphics[width=0.49\textwidth]{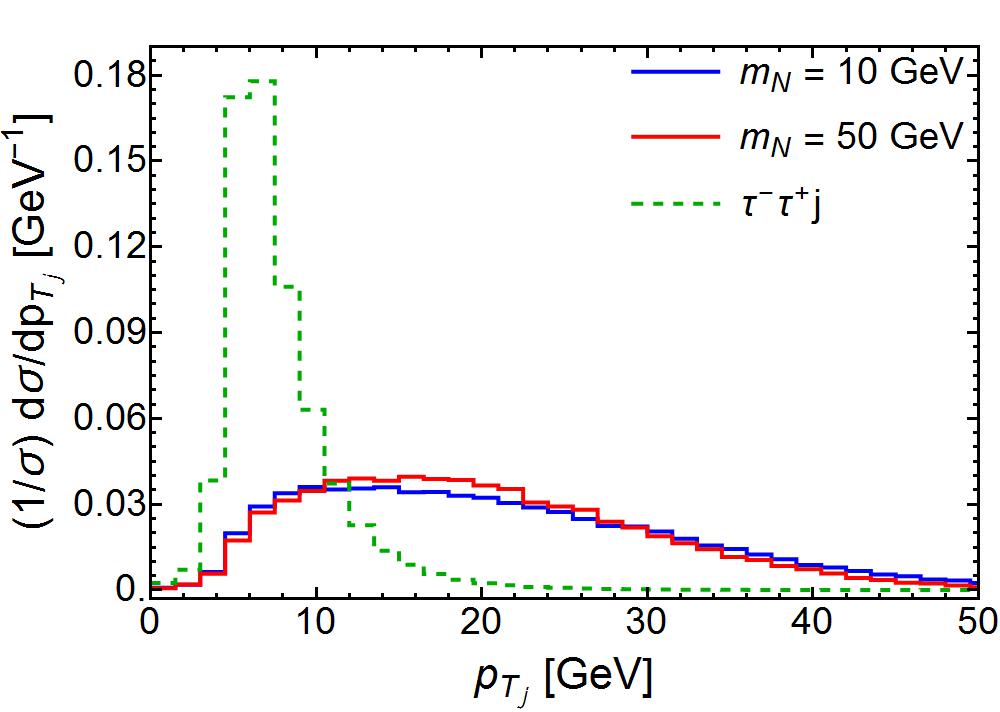}
		\caption{Transverse momentum distributions of leptons and the jet for signals with $m_N=10$ and 50 GeV and the $\tau^+ \tau^- j$ background in the $e^+ \mu^- j + E_T^{\text{miss}}$ final state, before applying the cuts of Eqs.~(\ref{eq:lepcut2})$-$(\ref{eq:drlj2}).}
		\label{fig:dist_e+mu-j_pre_lepton}
	\end{figure}
	
	Note that in this analysis we select leptons with $|\eta_{\ell}| < 3.5$ compared to the $e^+ + 3j$ analysis, where we select positrons with only positive $\eta$ values. The skewness of signal leptons in the forward hemisphere is not as prominent in this analysis, and hence less discriminating, as opposed to the previous analysis. However, $\eta_j$ does possess some discriminating power but we have chosen not to impose a cut on this variable for the purpose of optimization of signal significance. The transverse momentum distributions of $\ell_{1,2}$ (top panel), and $j$ (bottom panel) are shown in Fig.~\ref{fig:dist_e+mu-j_pre_lepton}. In Table~\ref{tab:cut-flow_Majorana} we present the cut-flow table for the background as well as the signal for two representative HNL masses, $m_N= 10$ GeV and 50 GeV, with $|U_e|^2=1$. The basic cuts reduce the background by two orders of magnitude, but the signal reduction is only a factor of 6~(4) for $m_N= 10\, (50)$ GeV. Thereafter, we require one lepton to be a positron, and the other to be a negatively charged muon.
	
\begin{figure}
		\centering
		\includegraphics[width=0.49\textwidth]{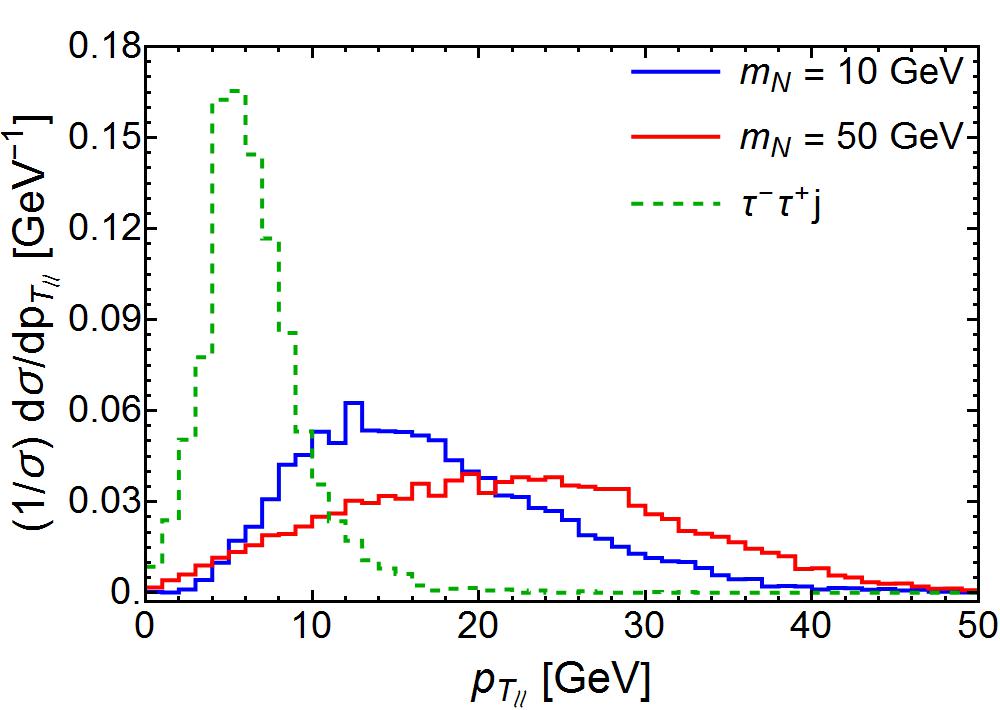}
		\includegraphics[width=0.48\textwidth]{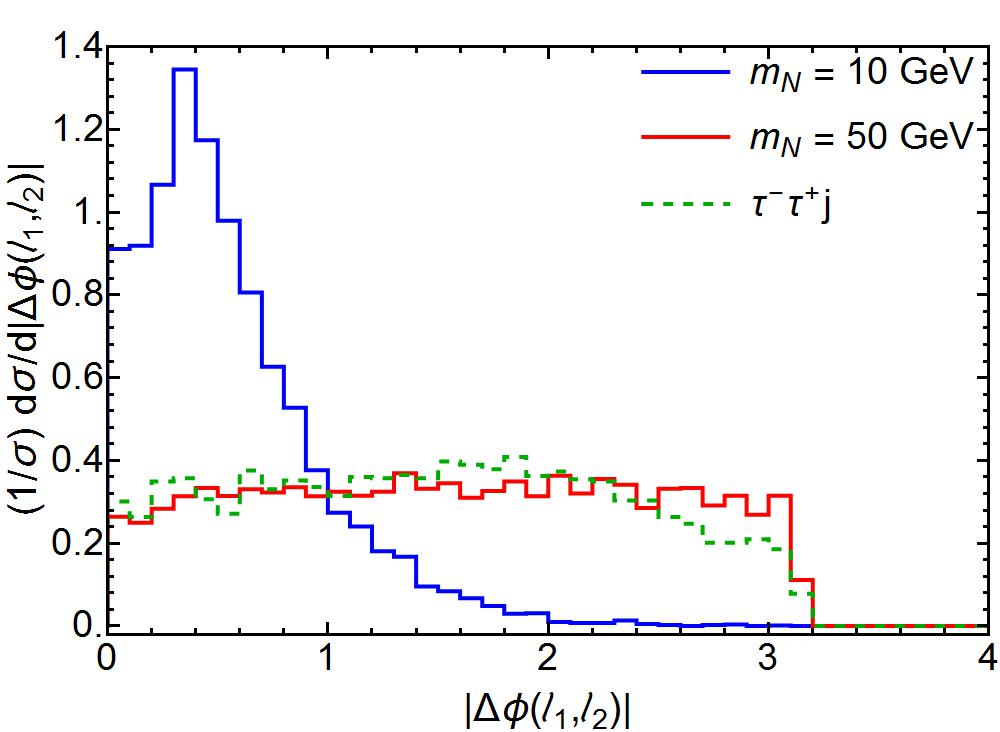}
		\includegraphics[width=0.49\textwidth]{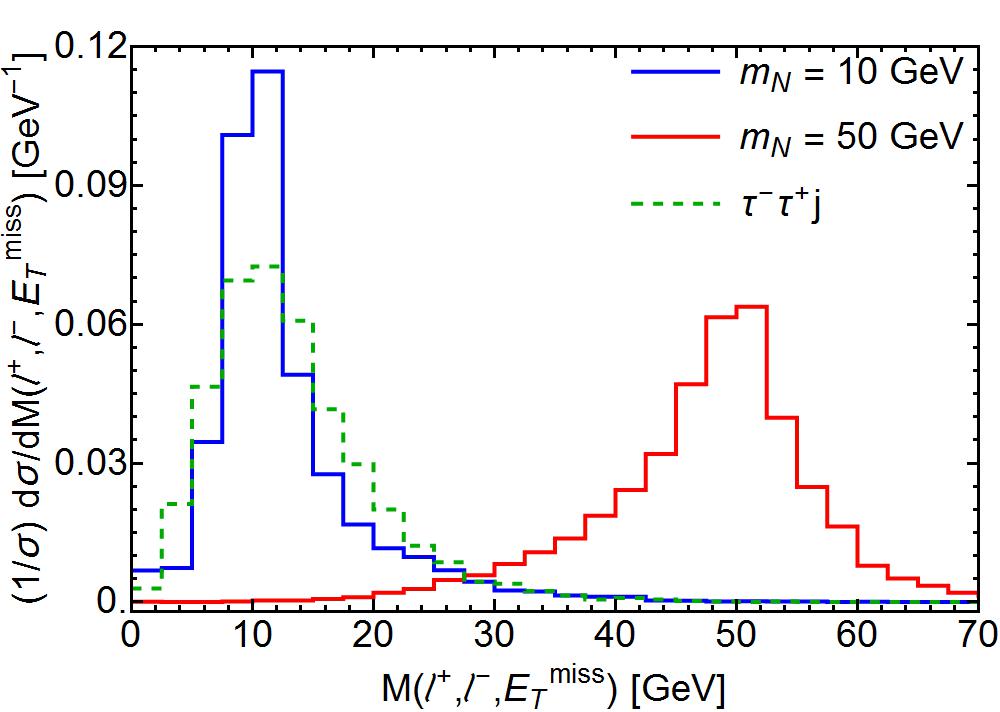}
		\caption{Kinematic distributions of $p_{T_{\ell \ell}}$, $|\Delta \phi (\ell_1, \ell_2)|$, and $M(\ell^+, \ell^-, E_T^{\text{miss}})$ for signals with $m_N=10$ and 50 GeV and the $\tau^+ \tau^- j$ background in the $e^+ \mu^- j + E_T^{\text{miss}}$ final state, before applying the $p_{T_{\ell \ell}}$ cut in Table~\ref{tab:cut-flow_Majorana}.}
		\label{fig:dist_e+mu-j_pre_pTll}
	\end{figure}

	Since the tau leptons in the background are predominantly produced from soft photons radiated by incoming electrons, the transverse momentum of the di-lepton system arising from tau decays peaks towards small values and falls sharply. In contrast, the same quantity for the signal has a much longer tail, as shown in the top left panel of Fig.~\ref{fig:dist_e+mu-j_pre_pTll}.
	Therefore, we impose the cut $p_{T_{\ell \ell}} > 12$ GeV, which suppresses the background by another order of magnitude while sacrificing only $\sim 20-30 \%$ of the signal events. 
	
	The cuts outlined above already lead to good sensitivities for $m_N < 20$ GeV. Nevertheless, we apply another cut for lighter HNLs as well to optimize their statistical significance a bit further. On the other hand, for heavier HNLs, another stringent cut is needed to suppress the background considerably to obtain good sensitivities. From this stage we use different cuts for signals with $m_N < 20$ GeV and $m_N\geq 20~\GeV$.

	For lighter HNL cases we employ $\Delta \phi (\ell_1, \ell_2) < 1$, which lowers the background by another factor of 2 while keeping the signal essentially intact. This is due to the fact that lighter HNLs produced at the EIC, are fairly boosted, leading to a small opening angle between leptons coming from HNL decays. In contrast, the $\Delta \phi (\ell_1, \ell_2)$ distribution for the background is flat. We present the $\Delta \phi (\ell_1, \ell_2)$ distributions of signal and background events in the top right panel of Fig.~\ref{fig:dist_e+mu-j_pre_pTll}.

	To improve heavier HNL sensitivities for the EIC,  we notice from the left panel of Fig.~\ref{fig:BrN} that HNL decay widths are quite small. Consequently, the absolute difference between the invariant mass of the di-lepton + $E_T^{\text{miss}}$ system and the HNL mass exhibits a narrow peak but only a small amount of background is contained within that peak. So, we apply the cut  $|M(\ell^+, \ell^-, E_T^{\text{miss}}) - m_N| <$ 10 GeV to diminish the SM background by another two orders of magnitude but losing only $20 \%$ signal for our representative $m_N=50$ GeV point as illustrated in the bottom panel of Fig.~\ref{fig:dist_e+mu-j_pre_pTll}. One can also infer from that plot that this cut will not work for lower HNL masses as the background also peaks around 10 GeV. For $m_N \geq 70$ GeV, a wider window cut of 30 GeV is used to accept more signal events as the $\tau^+\tau^- j$ background is almost negligible for $M(\ell^+, \ell^-, E_T^{\text{miss}}) > 40$ GeV. 
	
	One can perhaps improve the HNL sensitivity at the EIC by combining all four $\ell^+ \ell^- \, + \, E_T^{\text{miss}}$ channels. However, for the three non-LNV channels, $\ell^+ \ell^- \nu j$ and $\ell^+ \ell^- j$ give rise to significant backgrounds. So, even if one can improve the significance the $S/B$ ratio will decrease markedly and is thus susceptible to significant dilution in sensitivity in the presence of large systematic uncertainties. Since the EIC is at its early stages of design, it is {premature to reliably estimate these potential systematic uncertainties for} the backgrounds. 
	Therefore, we adopt a conservative approach and consider only the LNV channel. In this channel the SM background is negligible and our conclusions will be more robust against the possibility of large systematics at the EIC.
	
	\begin{table}[tb]
		\begin{center}
			\resizebox{\columnwidth}{!}{
				\begin{tabular}{|c|c c| c|}
					\hline
					\multirow{3}{*}{ Cut selection} & \multicolumn{2}{c|}{Signal $[e^- p \to (N\to\ell^-\ell^+\nu)j]$} & $\tau^-\tau^+ j \rightarrow$\\
					\cline{2-3}
					& $m_N = 10$ GeV & $m_N = 50$ GeV &  $ \ell^- \ell^+ j + 4\nu$\\ 
					& [pb] & [pb] & [pb] \\
					\hline   
					\hline
					Production & 3.16 & 0.55 & $0.05$\\
					\hline 
					Exactly $2 \ell$: &  \multirow{2}{*}{2.10} & \multirow{2}{*}{0.53} & \multirow{2}{*}{$0.01$}\\
					$p_{T_{\ell_{1,2}}} > 2$ GeV, $|\eta_{\ell_{1,2}}| < 3.5$ & & &\\
					\hline
					Exactly $1 j$: &  \multirow{2}{*}{1.82} & \multirow{2}{*}{0.44} & \multirow{2}{*}{$3.19 \times 10^{-3}$}\\
					{ $p_{T_j} > 10$ GeV}, $|\eta_{j}| < 3.5$ & & & \\
					\hline
					Isolation: &  \multirow{2}{*}{1.61} & \multirow{2}{*}{0.43} & \multirow{2}{*}{$3.13 \times 10^{-3}$}\\
					{$\Delta R (\ell_1,\ell_2) > 0.3$}, $\Delta R (\ell_{1,2},j) > 0.4$  
					& & &\\
					\hline
					\hline
					Require one $\mu^-$ and one $e^+$ & 0.51 & 0.13 & $7.83 \times 10^{-4}$\\
					\hline
					$p_{T_{\ell \ell}} > 12$ GeV & 0.37 & 0.10 & $3.90 \times 10^{-5}$\\
					\hline
					$|\Delta \phi (\ell_1, \ell_2)| < 1$ [$m_N < 20$ GeV]& 0.35 & $\times$ & $1.72 \times 10^{-5}$\\  
					\hline 
					$|M(\ell^+, \ell^-, E_T^{\text{miss}}) - m_N| < 10$ GeV [$m_N \geq 20$ GeV] & $\times$ & 0.08 & $2.07 \times 10^{-7}$\\  
					\hline 
					Polarization $P_e=-70\%$ & $\times1.7$  & $\times1.7$  &  $\times1$\\
					\hline
				\end{tabular}
			}
		\end{center}
		\caption{\label{tab:cut-flow_Majorana} Cut-flow table of the Majorana HNL signal, with $|U_e|^2=1$ in the $\mu^- e^+ j + E_T^{\text{miss}}$ final state. 
		}
	\end{table}
	
	\subsubsection{$\ell^- \ell^+ j \, + \, E_T^{\text{miss}}$ search for Dirac HNL}
	We now shift our attention to the prospects of finding a Dirac HNL at the EIC using the $\ell^- \ell^+ j \, + \, E_T^{\text{miss}} \, (\ell = e, \mu)$ channel. For the Dirac HNL we no longer have the LNV final state at our disposal. Therefore, we consider all three lepton-number-conserving di-lepton final states in this analysis.\footnote{{We note that, although those channels are unique for a Dirac HNL, a Majorana state will also contribute to the lepton-number conserving mode equally.}} Hence, we have to consider $\ell^+ \ell^- \nu j$ and $\ell^+ \ell^- j$ backgrounds on top of $\tau^+ \tau^- j \rightarrow \ell^+ \ell^- j + 4\nu$.

	We use the same selection and isolation criteria for the leptons and the solitary jet as described in Subsection~\ref{sec:prompt_emu_Majorana}. In Table~\ref{tab:cut-flow_Dirac} we show the cut-flow table for three representative HNL masses, $m_N= 5$, 10 and 50 GeV, with $|U_e|^2=1$. Hereby, for this channel our strategy to search for a Dirac HNL differs at places from the analysis presented in the previous subsection for a Majorana HNL. This is because in this study two extra backgrounds are involved, including the irreducible $\ell^+ \ell^- \nu j$, which respond to many cuts differently from the $\tau^+ \tau^- j$ background hitherto considered. 
	
	\begin{table}[tb]
		\begin{center}
			\resizebox{\columnwidth}{!}{%
				\begin{tabular}{|c|c c c| c c c|}
					\hline
					\multirow{2}{*}{ Cut selection} & \multicolumn{3}{c|}{Signal $[e^- p \to (N\to \ell^+ \ell^- \nu)j]$} & $\ell^+ \ell^- \nu_{\ell} \,j$ & $\ell^+ \ell^- j$ &  $\tau^-\tau^+ j \rightarrow$\\
					&$m_N = 5$ GeV &  $m_N = 10$ GeV & $m_N = 50$ GeV &  & & $ \ell^- \ell^+ j + 4\nu$\\ 
					& [pb] & [pb] & [pb] & [pb] & [pb] & [pb]\\
					\hline   
					\hline
					Production & 3.98 & 3.38 & 0.55 & $2.20 \times 10^{-3}$ & 5.06 & 0.05 \\
					\hline 
					Exactly $2 \ell$: & \multirow{2}{*}{2.05} & \multirow{2}{*}{1.95} &\multirow{2}{*}{0.53} & \multirow{2}{*}{$9.68 \times 10^{-4}$} &  \multirow{2}{*}{2.65} & \multirow{2}{*}{0.01} \\
					$p_{T_{\ell_{1,2}}} > 2$ GeV, $|\eta_{\ell_{1,2}}| < 3.5$ & & & & & & \\
					\hline
					Exactly $1 j$: & \multirow{2}{*}{1.86} & \multirow{2}{*}{1.71} &\multirow{2}{*}{0.44} & \multirow{2}{*}{$7.48 \times 10^{-4}$} &  \multirow{2}{*}{0.35} & \multirow{2}{*}{$3.20 \times 10^{-3}$}\\
					$p_{T_j} > 10$ GeV, $|\eta_{j}| < 3.5$ & & & & & &\\
					\hline
					Isolation: & \multirow{2}{*}{1.25} & \multirow{2}{*}{1.58} & \multirow{2}{*}{0.43} & \multirow{2}{*}{$5.45 \times 10^{-4}$} &  \multirow{2}{*}{0.33} & \multirow{2}{*}{$3.14 \times 10^{-3}$}\\
					{$\Delta R (\ell_1,\ell_2) > 0.3$}, $\Delta R (\ell_{1,2},j) > 0.4$  
					& & & & & &\\
					\hline
					\hline
					$E_T^{\text{miss}} > 5$ GeV & 0.80 & 1.07 & 0.40 & $5.32 \times 10^{-4}$ & 0.02  & $2.46 \times 10^{-3}$\\
					\hline 
					$p_{T_{\ell \ell}} > 12$ GeV & 0.43 & 0.64 & 0.29 & $1.50 \times 10^{-4}$ & $5.47 \times 10^{-3}$ & $8.90 \times 10^{-5}$ \\
					\hline 
					\multirow{3}{*}{$|M(\ell^+, \ell^-, E_T^{\text{miss}}) - m_N| < 5$ GeV} & 0.27 & $\times$ & $\times$ & $2.39 \times 10^{-6}$ & $5.97 \times 10^{-4}$ & $1.56 \times 10^{-5}$\\
					& $\times$ & 0.42 & $\times$ & $7.12 \times 10^{-6}$ & $1.37 \times 10^{-3}$ & $3.15 \times 10^{-5}$ \\
					& $\times$ & $\times$ & 0.17 & $2.34 \times 10^{-5}$ & $1.42 \times 10^{-4}$ & $4.15 \times 10^{-7}$ \\
					\hline
					$M(\ell^+ \ell^- j) > 45$ GeV [$m_N < 10$ GeV] & 0.18 & $\times$ & $\times$ & $1.34 \times 10^{-6}$ & $1.82 \times 10^{-4}$ & $6.43 \times 10^{-6}$\\
					\hline
					\multirow{2}{*}{$0.2 < |\Delta \phi (j, E_T^{\text{miss}})| < 3$ [$m_N \geq 10$ GeV]}  & $\times$ & 0.24 & $\times$  & $5.00 \times 10^{-6}$ & -- & $9.75 \times 10^{-6}$\\
					& $\times$ & $\times$ & 0.16 & $2.06 \times 10^{-5}$ & -- & $2.07 \times 10^{-7}$ \\
					\hline
					Polarization $P_e=-70\%$ & $\times1.7$ & $\times1.7$ & $\times1.7$ & $\times1.6$ & $\times1$ & $\times1$ \\
					\hline
				\end{tabular}
			}
		\end{center}
		\caption{\label{tab:cut-flow_Dirac} Cut-flow table of the Dirac HNL signal, with $|U_e|^2=1$, and SM backgrounds in the $\ell^- \ell^+ j + E_T^{\text{miss}}$ final state. The ``--" indicates the background size is negligible.
		}
	\end{table}
	
	In the previous subsection we presented the cut-flows for $m_N= 10$ and 50 GeV only as we used two different strategies for $m_N < 20$ GeV and $m_N \geq 20$ GeV. To be consistent with other prompt analyses presented in this paper, we still show cut-flows for $m_N= 10$ and 50 GeV. However, we use the same strategy for both these cases in this search, but use a different one for $m_N < 10$ GeV cases. Hence, we additionally show the cut-flow table for $m_N = 5$ GeV in Table~\ref{tab:cut-flow_Dirac}.
	
	\begin{figure}
		\centering
		\includegraphics[width=0.49\textwidth]{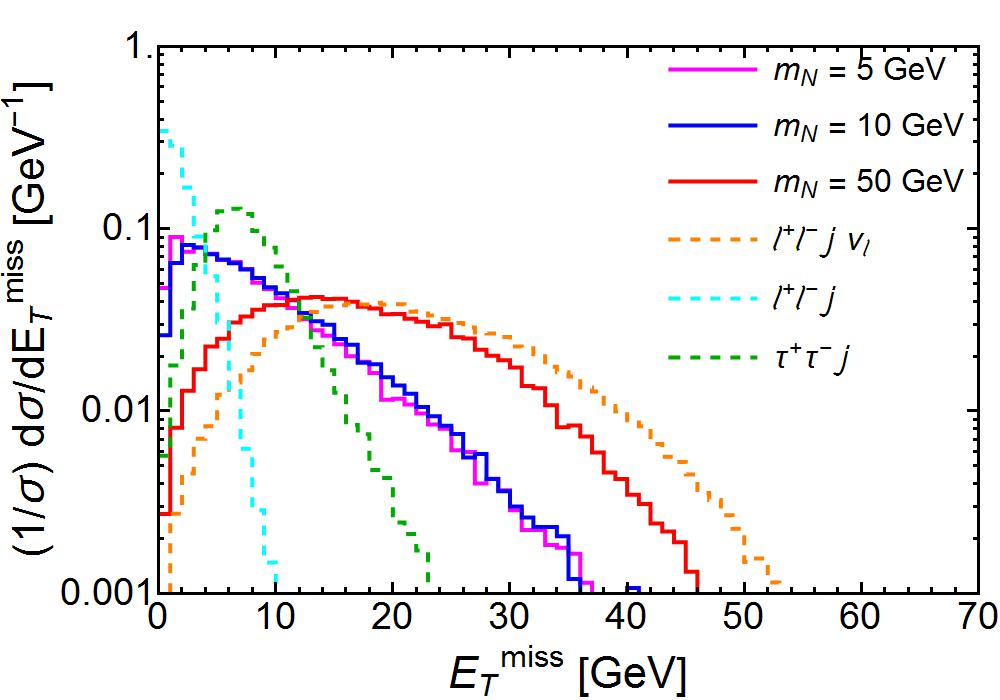}
		\includegraphics[width=0.48\textwidth]{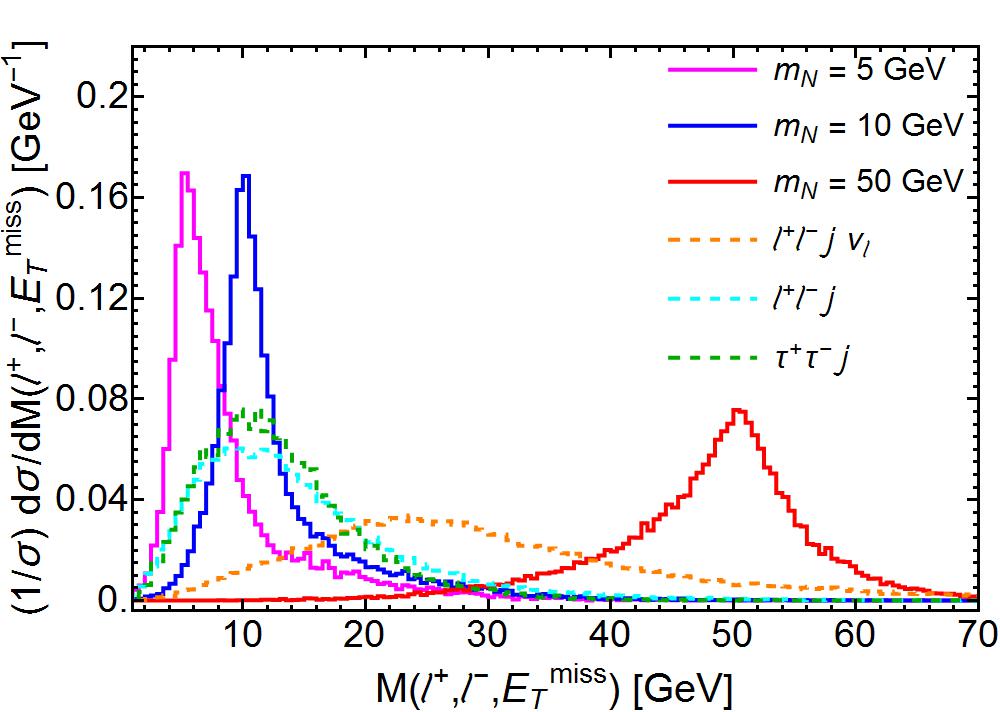}
		\includegraphics[width=0.49\textwidth]{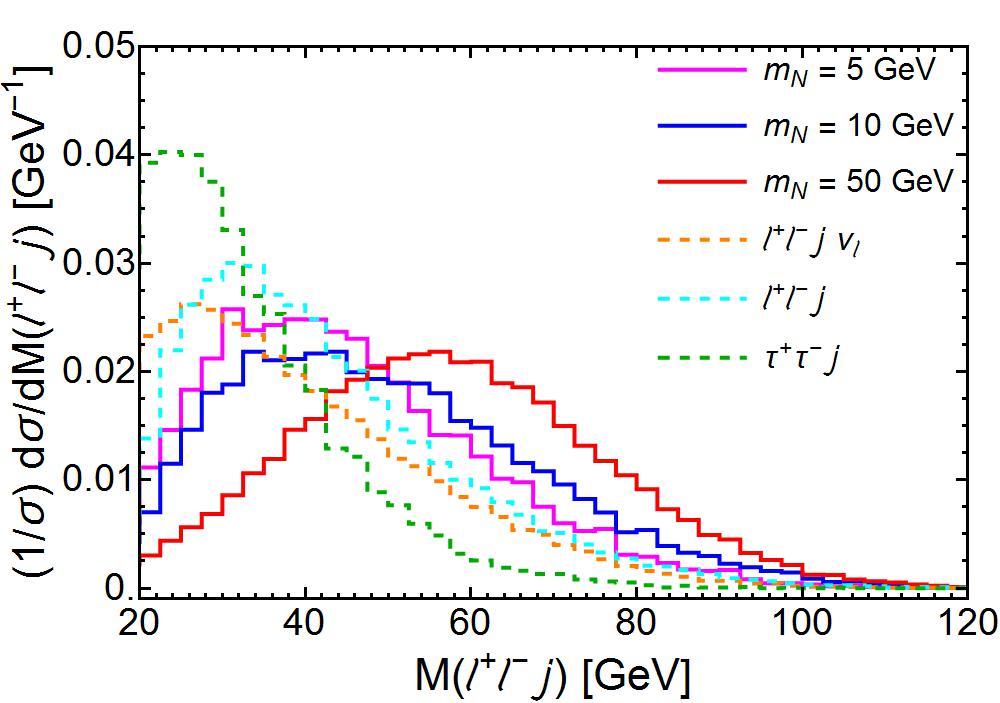}
		\includegraphics[width=0.48\textwidth]{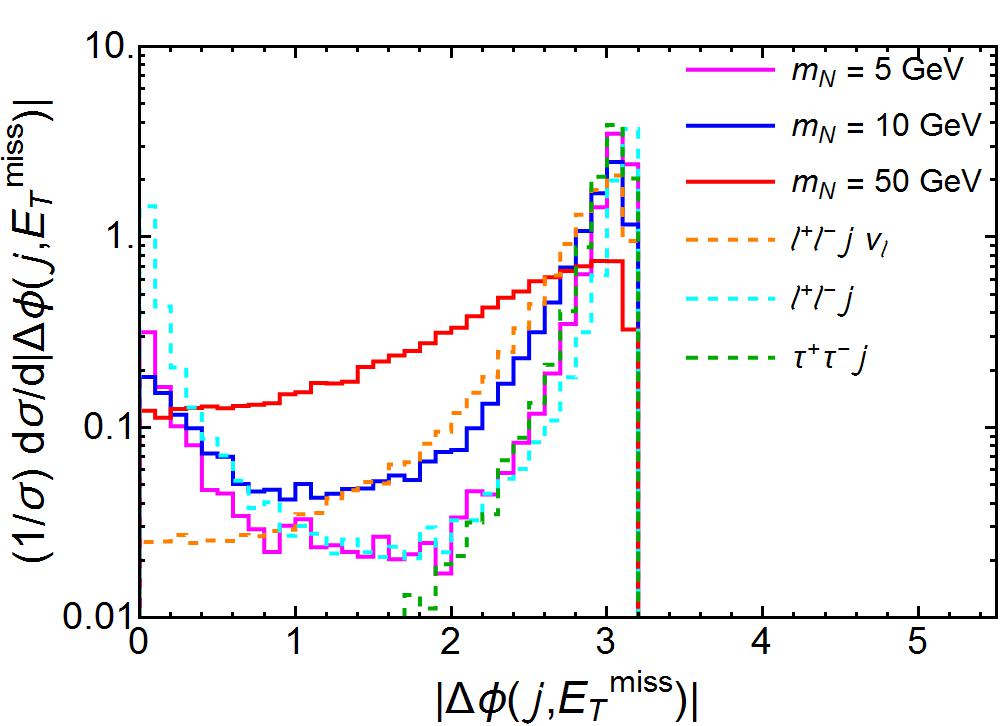}
		\caption{Kinematic distributions of $E_T^{\text{miss}}$ (top left), $M(\ell^+, \ell^-, E_T^{\text{miss}})$ (top right), $M_{\ell \ell j}$ (bottom left), and $|\Delta \phi (j, E_T^{\text{miss}})|$ (bottom right) for signals with $m_N=5$, 10 and 50 GeV and the three SM backgrounds in the $\ell^+ \ell^- j + E_T^{\text{miss}} \, (\ell = e, \mu)$ final state, after applying the isolation cuts of Eq.~(\ref{eq:drlj2}).}
		\label{fig:dist_l+l-j_post_isolation}
	\end{figure}

	After the selection of isolated leptons and the jet, we demand the events to have some missing transverse momentum ($E_T^{\text{miss}}$). With the current detector design with the far-forward coverage for the electrons and hadrons \cite{EIC-detector-handbook}, one expects to achieve high granularity for good $E_T^{\text{miss}}$ determination.
	We use a nominal missing energy cut of 5 GeV.
	This cut is used to suppress the $\ell^+ \ell^- j$ background, where the source of $E_T^{\text{miss}}$ comes from jet energy mis-measurement and is expected to peak at very small values as can be seen from the top left panel of Fig.~\ref{fig:dist_l+l-j_post_isolation}. This cut reduces the $\ell^+ \ell^- j$ by an order of magnitude without significant loss of signal events.

	Next, we impose $p_{T_{\ell \ell}} > 12$ GeV and $|M(\ell^+, \ell^-, E_T^{\text{miss}}) - m_N| < 5$ GeV cuts successively. Again, the $|M(\ell^+, \ell^-, E_T^{\text{miss}}) - m_N|$ cut is sensitive to $m_N$ and leads to different background efficiencies for different $m_N$ values. The reader may recall that we argued against using the above invariant mass window cut for $m_N \leq 10$ GeV in the $e^+ \mu^- j + E_T^{\text{miss}}$ study since the $M(\ell^+, \ell^-, E_T^{\text{miss}})$ distribution for $\tau^+ \tau^- j$ peaks around 10 GeV. Nonetheless, we apply this cut in the present case of Dirac HNLs as we have to deal with a far more problematic background -- the irreducible $\ell^+ \ell^- \nu j$, which the window cut brings down significantly. Collectively, for $m_N=5$ and 10 GeV cases, the two cuts above reduce the $\ell^+ \ell^- \nu j$ background by two orders of magnitude and the other two backgrounds by an order of magnitude each. In contrast, for the $m_N=50$ GeV case, the three backgrounds are suppressed by one, two, and four orders of magnitude, respectively. These cuts retain $\sim 60 \%$ of signal events in all three cases. The $M(\ell^+, \ell^-, E_T^{\text{miss}})$ for all the signal benchmark points and backgrounds are presented in the top right panel of Fig.~\ref{fig:dist_l+l-j_post_isolation}. For $m_N \geq 70$ GeV, a wider window cut of 10 GeV is used.
	
	Thereafter, we employ separate cuts for the $m_N < 10 $ and $m_N \geq 10$ GeV scenarios. For lighter HNLs we use the cut $M(\ell^+ \ell^- j ) > 45$ GeV leading to a factor of $2-3$ suppression of all three backgrounds and improving the EIC sensitivities for light HNLs. The signal efficiency of the cut is $67\%$ for $m_N = 5$ GeV.
	
	For $m_N \geq 10$ GeV we achieve stronger background suppression by using the cut\\ $0.2 < |\Delta \phi (j, E_T^{\text{miss}})| < 3$. This cut renders the  $\ell^+ \ell^- j$ background negligible for our analysis of heavier Dirac HNLs. The other two backgrounds are already small. The signal efficiency of this cut improves with increasing $m_N$, from $~57\%$ for $m_N = 10$ GeV to almost $100\%$ for $m_N \sim 100$ GeV. We show the $M_{\ell \ell j}$ and $|\Delta \phi (j, E_T^{\text{miss}})|$ distributions for signal and background events in the bottom left and right panels of Fig.~\ref{fig:dist_l+l-j_post_isolation}, respectively. 
	
	
	\subsubsection{Summary of Prompt HNL Searches} 
	\label{sec:prompt-summary}
	
\begin{figure}[tb]
		\begin{center}
			\includegraphics[width=0.8\textwidth]{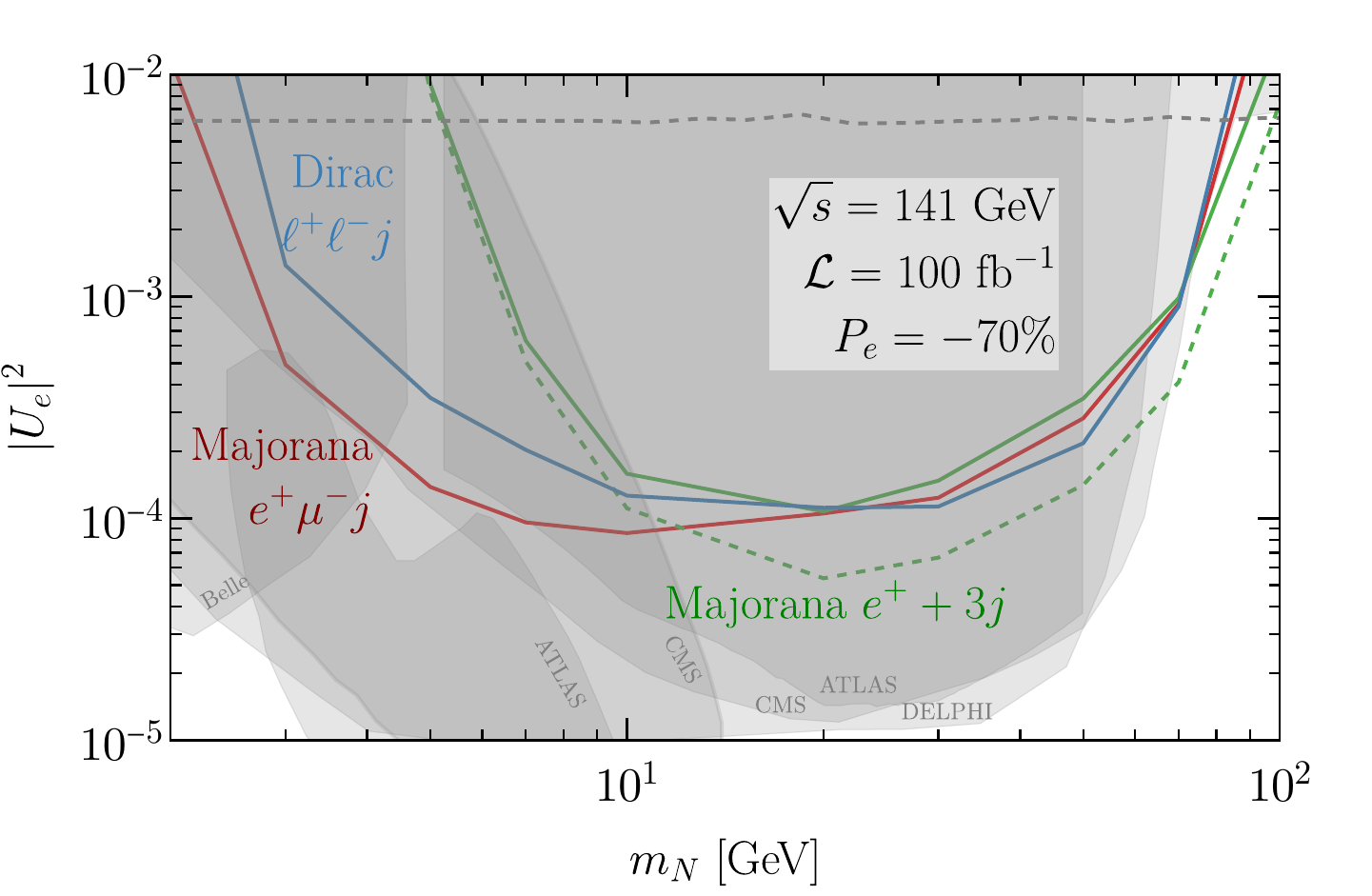}
		\end{center}
		\caption{The expected 95$\%$ C.L. exclusion limits from prompt searches at the EIC with $\sqrt{s}=141$ GeV and 100 fb$^{-1}$ of integrated luminosity for HNLs (colored lines), compared with the existing bounds from direct searches~\cite{DELPHI:1996qcc,Belle:2013ytx,CMS:2018iaf,CMS:2022fut,ATLAS:2019kpx,ATLAS:2022atq} (gray shaded regions) and indirect precision electroweak constraints~\cite{MuLan:2012sih} (horizontal dashed line).
			The solid (dashed) green line indicates the sensitivity of the prompt Majorana HNL decay $N\to e^++3j$, with a misidentification rate assumed as 0.1\% (0.01\%).}
		\label{fig:prompt}
	\end{figure}

	We summarize our results for the prompt searches obtained in this section in Fig.~\ref{fig:prompt}. We plot the $95\%$ C.L. exclusion curves determined from the above analyses by setting the metric $\mathcal{S}= S / \sqrt{S+B}=1.96$, where $S$ and $B$ are signal and background events, respectively, after all the cuts. We also compare them to existing direct bounds on HNLs from
	CHARM~\cite{CHARM:1985nku},
	DELPHI~\cite{DELPHI:1996qcc},
	Belle~\cite{Belle:2013ytx},
	CMS~\cite{CMS:2018iaf,CMS:2022fut},
	ATLAS~\cite{ATLAS:2019kpx,ATLAS:2022atq}
	experiments, as well as the indirect EW precision constraint from the MuLan data~\cite{MuLan:2012sih}.
	We note that the global EW fits~\cite{delAguila:2008pw,Akhmedov:2013hec,Basso:2013jka,deBlas:2013gla,Antusch:2015mia,Deppisch:2015qwa} can place slightly stronger indirect constraints than the MuLan bound, depending on various specific assumptions.
	We observe that the EIC can most sensitively probe $|U_e|^2$ for HNL masses between about 10 and 50 GeV. Above $m_N \sim 50$ GeV, the bounds relax fast due to rapidly falling production cross-sections. In contrast, for low HNL masses, the reach of the prompt searches is limited by the isolation criteria of leptons and jets. As already mentioned earlier, for smaller $m_N$ values, HNLs produced at the EIC are significantly boosted and the decay products are extremely collimated. In this regime, the 2 lepton searches perform better than the $e^+ + 3j$ search. This is because we defined an isolated lepton with an isolation cone ($\Delta R$) of 0.3 around it, while for the jets a value of 0.4 is used for the same. It is worth noting that for low masses and mixing angles the HNL decays can be displaced, leading to a reduction in the number of prompt signal events. While we have accounted for this effect,  we find that the isolation cuts provide the dominant limiting factor to the reach at low masses. Relaxing the isolation requirements would improve the signal acceptance in the low mass region, leading to an improvement in the reach. 
	
	It is also worth emphasizing that if one can improve the electron charge misidentification and achieve $f^{\text{MID}}=0.01 \%$, the EIC can impose limits on $|U_e|^2$ in the $e^+ + 3j$ channel for $70 < m_N < 90$ GeV, which is better than existing laboratory limits on $|U_e|^2$.  In this mass range the strongest existing bounds come from the CMS $3 \ell$+$E_T^{\text{miss}}$ analysis~\cite{CMS:2018iaf}. At the LHC, the main SM backgrounds for the HNL search are 
	$WZ, \, ZZ/\gamma^*$ and leptons coming from top quark and heavy meson cascade decays, and all these SM processes are copiously produced. In contrast, in the cleaner environment of the EIC, the primary background of our $e^+ + 3j$ analysis is fake in nature and can be efficiently suppressed by a low $f^{\text{MID}}$ leading to the EIC outperforming the LHC in this mass window.

	\subsection{Displaced HNL search}
\begin{figure}[tbh]
		\begin{center}
			\includegraphics[width=.54\textwidth]{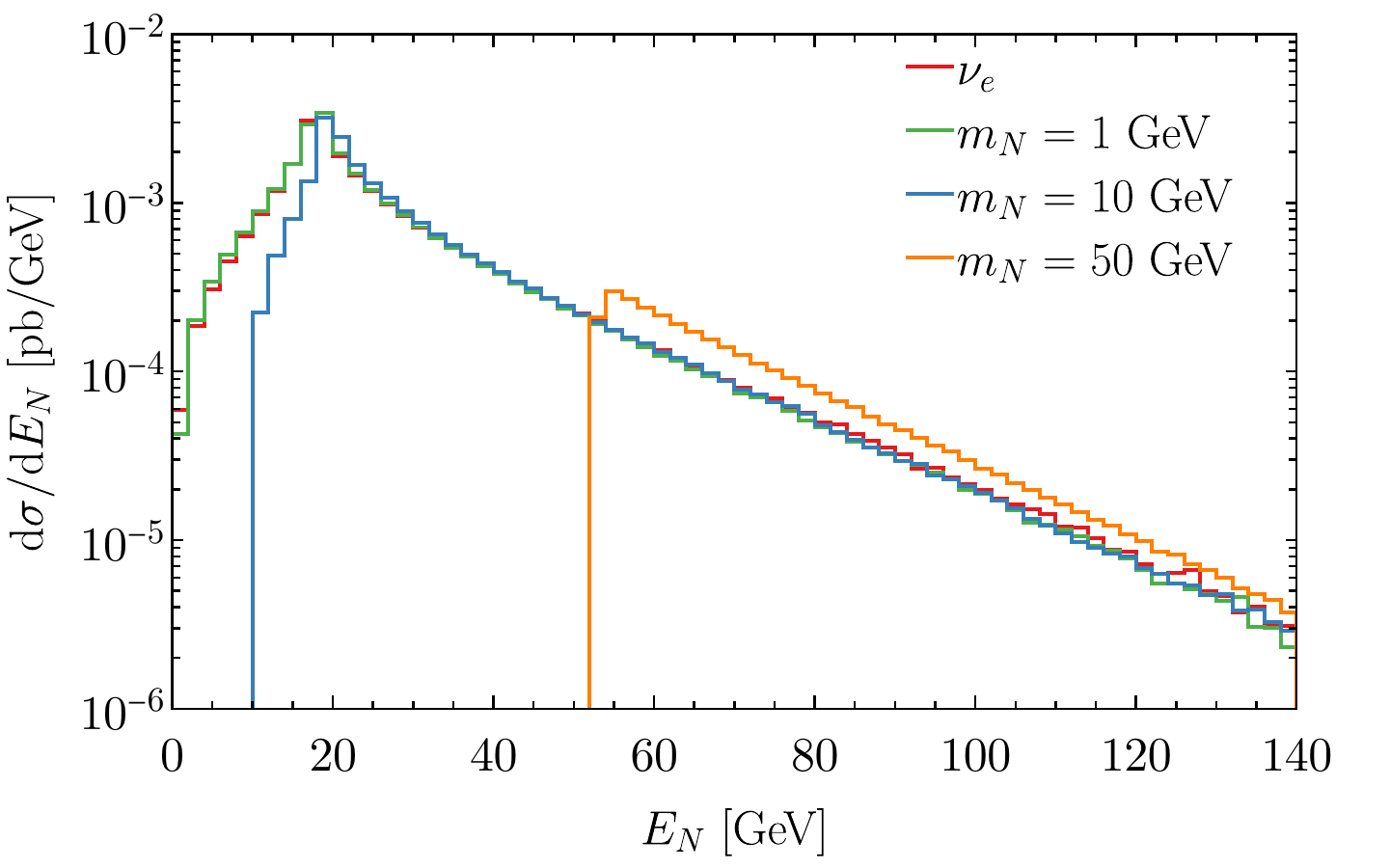}
			\includegraphics[width=.45\textwidth]{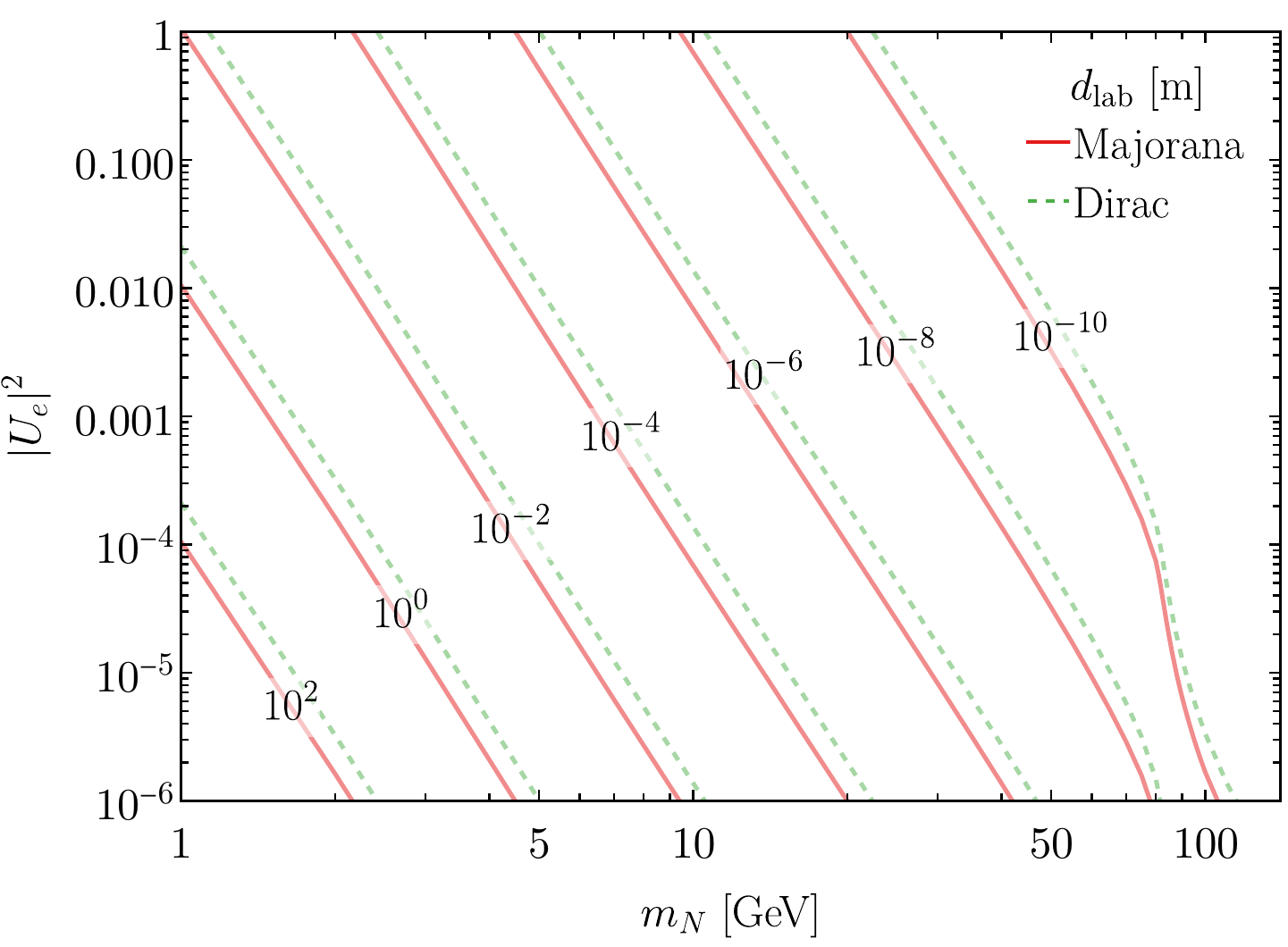}
		\end{center}
		\caption{Left: The energy distribution of $N$ ($\nu_e$) in the production channel $ep\to jN(\nu_e)$  
			at the EIC with beam energy as $E_p\times E_e=275\times18~\GeV^2$. 
			Right: 
			The typical decay length of HNL in the EIC lab frame estimated with the condition $E_N\sim\sqrt{E_e^2+m_N^2}$. 
		}
		\label{fig:DL}
	\end{figure}

	As shown in the left panel of Fig.~\ref{fig:BrN}, in the parameter space with a small mixing angle $U_e$ and small HNL mass $m_N$ the lifetime of $N$ becomes quite long, allowing it to travel macroscopic distances on the scale of the EIC detector before decaying. 
	In other words, the HNL behaves as a long-lived particle (LLP). 
	In the laboratory frame, the decay length of $N$ is given by 
	\begin{equation} \label{eq:DL}
		d_{\rm lab}=\gamma\beta c\tau_N,\quad \gamma=E_{N}/m_{N},
	\end{equation}
	which is determined by its proper lifetime $\tau_N$ in Eq.~(\ref{eq:tauN}) and its lab energy $E_N$. In Fig.~\ref{fig:DL}~(left), we show the energy distribution of HNL produced via $ep\to Nj$ at the EIC with beam energy as $275\times18~\GeV^2$. When $m_N\ll E_e$, the HNL energy is populated around $E_N\sim E_e$ due to a Jacobian peak. When $m_N\gg E_e$, HNLs are mainly produced around the threshold region with $E_N\sim m_N$. We can take an approximation $E_N\sim\sqrt{E_e^2+m_N^2}$ to smoothly bridge these two regions, which describes the energy peaks very well as shown in Fig.~\ref{fig:DL}.
	With this condition, we can estimate the characteristic decay length of the HNL in the EIC detector frame, which is shown in the $m_N-|U_e|^2$ plane in Fig.~\ref{fig:DL} (right). 
	When $m_N>E_e$, the lab decay length in $d_{\rm lab}$ is smaller than the proper one $c\tau$ estimated in Fig.~\ref{fig:BrN}, due to the Lorentz boost factor $\gamma\beta\sim E_e/m_N<1$. In contrast, when $m_N\ll E_e$, $d_{\rm lab}$ becomes significantly larger than $c\tau$. As we see in the small $|U_e|^2$ and small $m_N$ region, $d_{\rm lab}$ could range from sub $\mu$m to 100 m, which can be longer than those of heavy mesons such as $B^{0,\pm}, D^{0,\pm}$ and the $\tau$ lepton by several orders of magnitudes, offering the prospect of a low background search.
	
	Studies for long-lived HNLs have been performed for the LHC, demonstrating strong sensitivity to the HNL parameter space in the small $|U_e|^2$ and $m_N$ region; see \emph{e.g.}, Refs.~\cite{Drewes:2019fou,Liu:2019ayx} for phenomenological studies and Refs.~\cite{CMS:2022fut,ATLAS:2022atq} for experimental searches.
	Here, we focus on the displaced lepton signature of the long-lived HNL at the future EIC. 
	The representative Feynman diagrams are shown in Fig.~\ref{fig:feyn}. 
	In the charged-current decay case, Fig.~\ref{fig:feyn} (left), the track of the final-state electron in $N\to eW^*$ decay can provide evidence of the displaced HNL, distinguished from the prompt $ep$ primary vertex. 
	The virtual $W^*$ decay can go through either leptonic channel $W^*\to\ell\nu$ or the hadronic one $W^*\to q\bar{q}'$, of which both contribute to displaced signal events. In the neutral-current decay case, Fig.~\ref{fig:feyn} (right), we require at least one lepton in the final state, which can be only through $N\to\nu_e(Z^*\to\ell^+\ell^-)$, where $\ell=e,\mu$. We do not consider the $Z^*\to\tau^+\tau^-$ channel in this analysis, as the final-state $\tau$ lepton has different signatures and also suffers from smaller efficiencies and larger uncertainties in reconstruction. 
	
	Recently, two baseline concepts for the EIC tracking detectors have been discussed in Ref.~\cite{AbdulKhalek:2021gbh}. A fully realistic simulation of the displaced particle acceptance and detection capabilities of the proposed detectors goes beyond the scope of this work. Instead, we will consider a simplified analysis to estimate the potential reach of the EIC. We assume a cylindrical detector configuration representing the main tracker, with respective radius and length~\cite{AbdulKhalek:2021gbh}
	\begin{equation}
		r=0.4~{\rm m}, ~l=1.2~{\rm m}.
		\label{eq:cylinder}
	\end{equation}
	We require the HNL to decay within the cylinder and a displaced lepton $(e,\mu)$ with a nonzero transverse impact parameter $d_T$.   
	In our analysis, we will consider the following two conservative choices of $d_T$:
	\begin{equation}
		d_T=2~(20) \, {\rm mm}.
		\label{eq:dT}
	\end{equation}
	We note that the impact parameter cut (\ref{eq:dT}) is quite large compared to the estimated EIC tracking and vertexing resolution of order few $\mu$m~\cite{AbdulKhalek:2021gbh}. 
	While smaller impact parameter cuts could enhance the reach to shorter HNL lifetimes, these regions of parameter space are already constrained by past experiments. On the other hand, the large $d_T$ cut in Eq.~(\ref{eq:dT}) will significantly suppress SM heavy-flavor backgrounds. 
	
	To estimate the acceptance, we simulate $ep\to jN$ events, weighting each event according to the probability to 
	decay inside the cylinder, Eq.~(\ref{eq:cylinder}), with transverse displacement $l_T>d_T$ satisfying Eq.~(\ref{eq:dT}).
	Furthermore, to facilitate the reconstruction of displaced signal events, we impose the following basic acceptance cuts
	\begin{equation}\label{eq:DVScut}
		p_T^j>5~\GeV,~ p_T^\ell>2~\GeV,~|\eta_{j,\ell}|<3.5.
	\end{equation}
	Here, the jet cuts are designed to resolve the primary vertex, 
	while the lepton cuts ensure that the displaced lepton is easily detected.
	The lepton $p_T$ cut is motivated by the expected energy resolution of the electromagnetic calorimater~\cite{AbdulKhalek:2021gbh},
	consistent with our choices in the prompt searches discussed above.
	The main background comes from the leptonic decays of boosted heavy-quark hadrons, such as $B(D)\to\ell X$.
	However, the impact parameter  $d_T=2~(20)$ mm selection is one (two) orders of magnitude larger than the proper heavy hadron decay length, 
	which should thus allow us to mitigate heavy-hadron backgrounds, based on our simplified simulation.

\begin{figure}[tbh]
		\centering
		\includegraphics[width=0.8\textwidth]{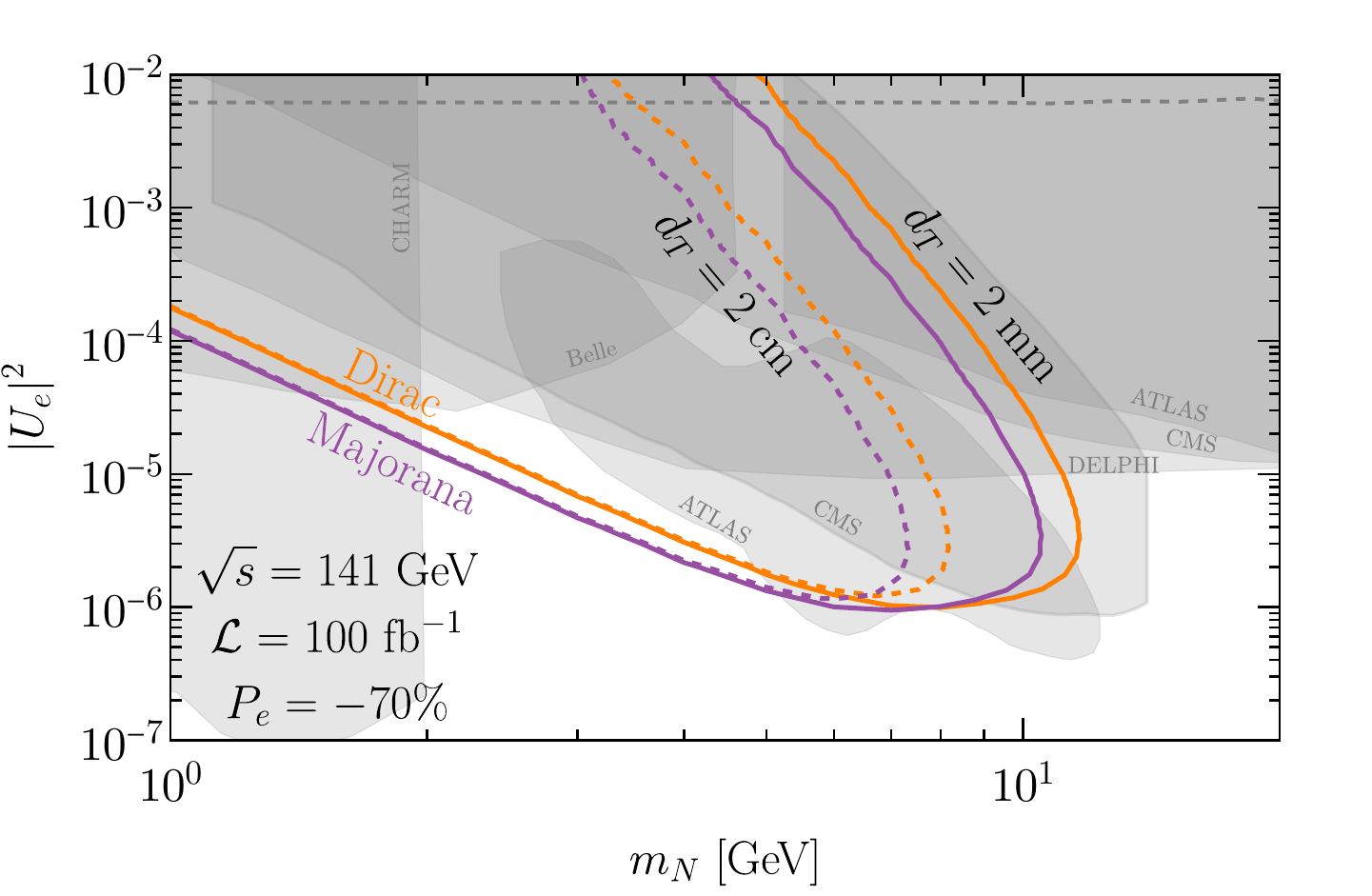}
		\caption{The expected contours of $N=5$ displaced vertex events detected in the EIC detector. The Majorana (Dirac) type events are shown as purple (orange) lines. The solid (dashes) lines indicate the impact parameter choice as $d_T=2~(20)$ mm. These results are compared with the existing bounds from direct searches~\cite{CHARM:1985nku,DELPHI:1996qcc,Belle:2013ytx,CMS:2018iaf,CMS:2022fut,ATLAS:2019kpx,ATLAS:2022atq} (gray shaded regions) and indirect precision electroweak constraints~\cite{MuLan:2012sih} (horizontal dashed line). In particular, we include existing displaced vertex searches in the 13 TeV CMS~\cite{CMS:2022fut} and ATLAS~\cite{ATLAS:2022atq} experiments (dark shaded islands).
		}
		\label{fig:DVS}
	\end{figure}

	In anticipation of a nearly background-free search, we show 5-event contours for both Dirac and Majorana HNLs in Fig.~\ref{fig:DVS}.
	This would correspond to a $95\%$ CL bound in the presence of one background event.
	The existing bounds from displaced HNL searches at CMS~\cite{CMS:2022fut} and ATLAS~\cite{ATLAS:2022atq} are shown as the dark-shaded islands in a similar mass -- mixing angle range. Our sensitivity curves display the characteristic features of an LLP search.
	The upper-right boundary is mainly driven by the impact parameter cut and the short decay length predicted for larger mixing angles and masses. 
	As mentioned above, a smaller cut on $d_T$ will extend the reach in this direction, but the parameter space is well covered already.
	The lower flat boundary is dictated by the signal event rate, as the production cross section scales as $\sigma\propto|U_e|^2.$
	Finally, the lower-left contours are determined by the tracker size, which is optimistically chosen as the distance of the most outside tracker disk $l=1.2$~m~\cite{AbdulKhalek:2021gbh}. We see that in comparison with the Dirac HNL, the Majorana type gives a better sensitivity in this direction, as a result of its larger decay width (smaller decay length).
	We observe that the EIC has the potential to cover new regions of parameter space in the GeV mass region with searches for displaced HNL decays.
	
	\subsection{Invisible decay search}
	
	\begin{figure}[tbh]
		\begin{center}
			\includegraphics[width=.45\textwidth]{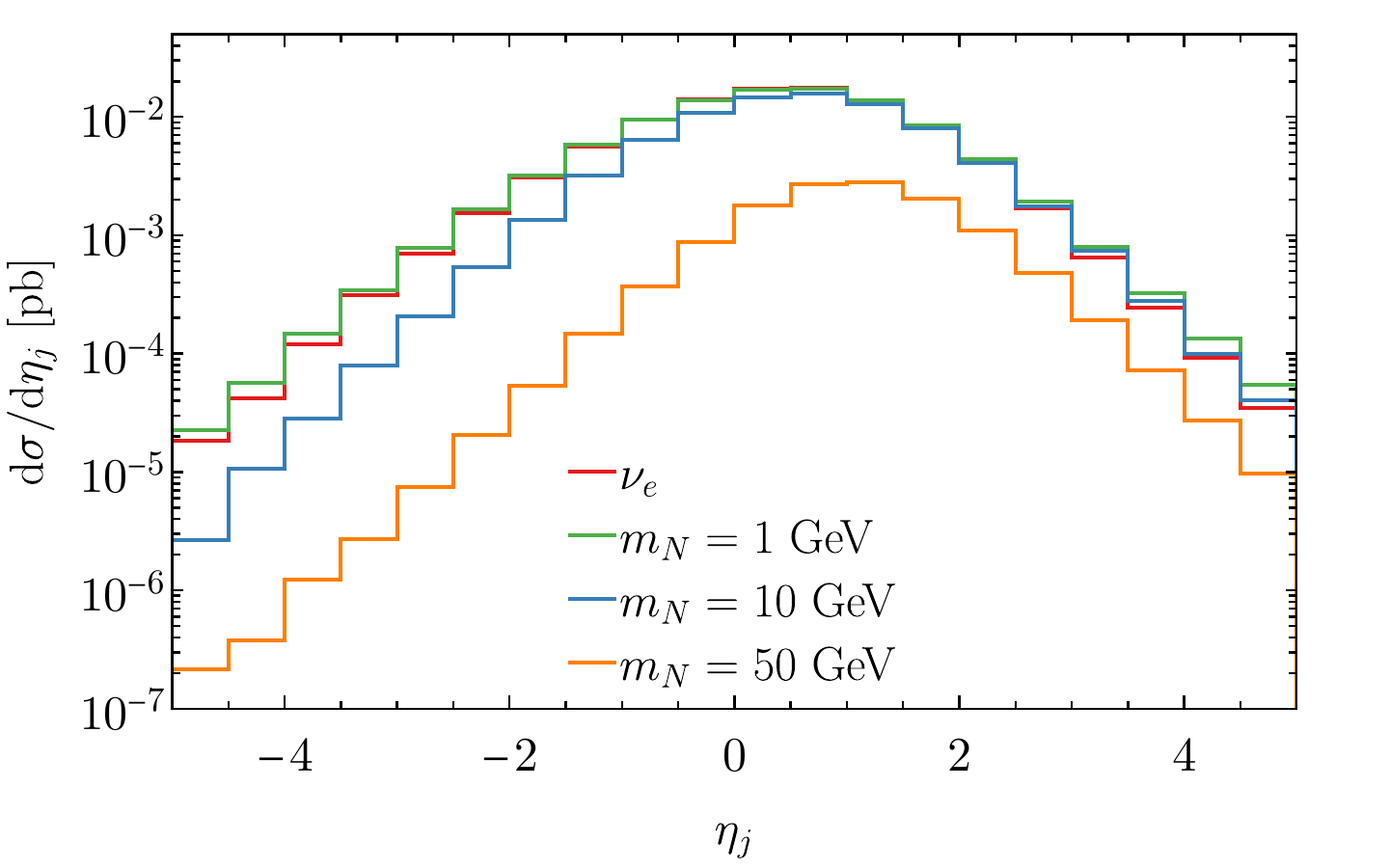}
			\includegraphics[width=.45\textwidth]{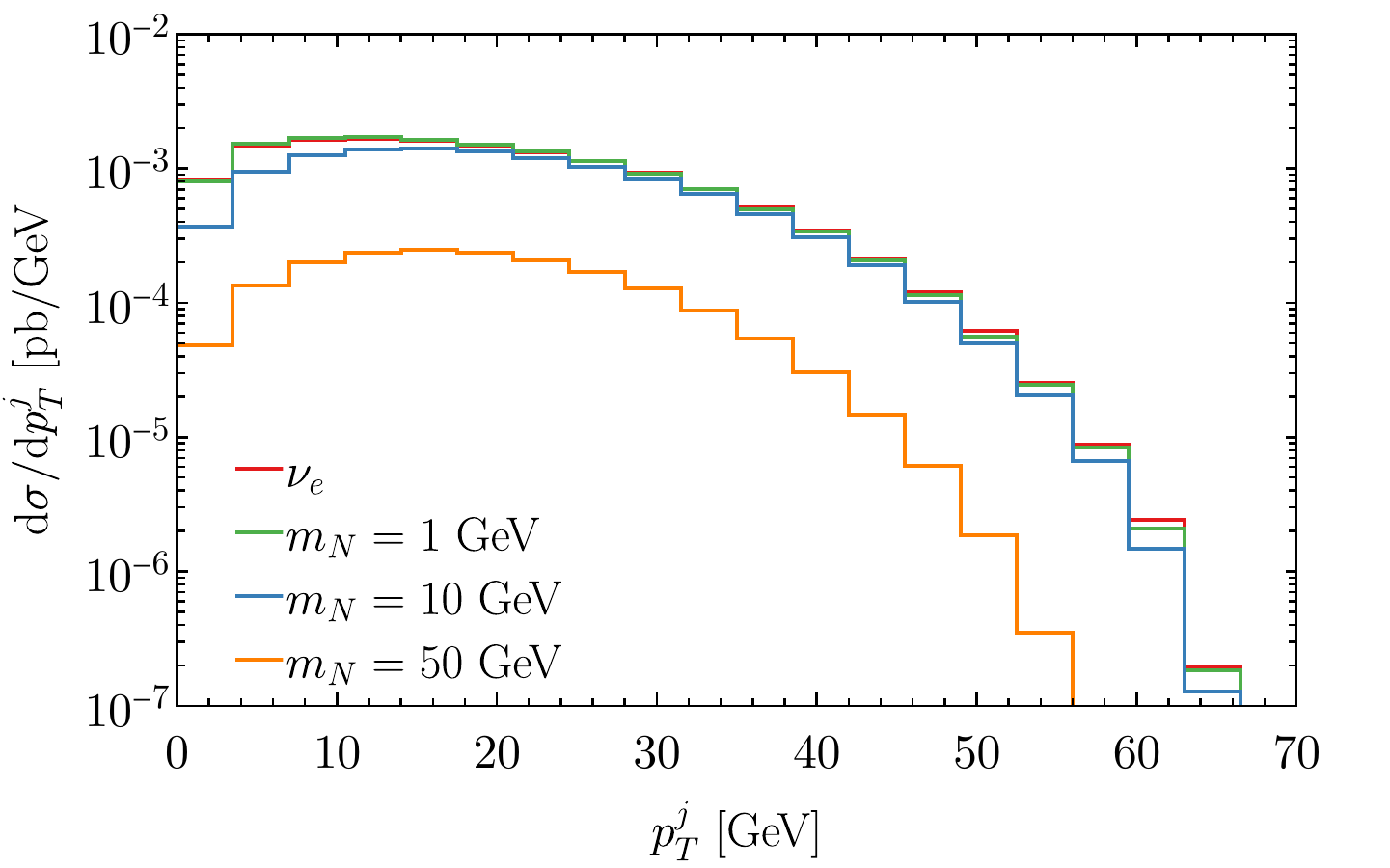}
		\end{center}
		\caption{Jet rapidity $\eta_j$ and transverse momentum $p_T^j$ distributions of mono-jet production at the EIC $\sqrt{s}=141~\GeV$, under the assumption $|U_e|^2=1$.}
		\label{fig:dist_monojet}
	\end{figure}
	
\begin{figure}
		\centering
		\includegraphics[width=0.5\textwidth]{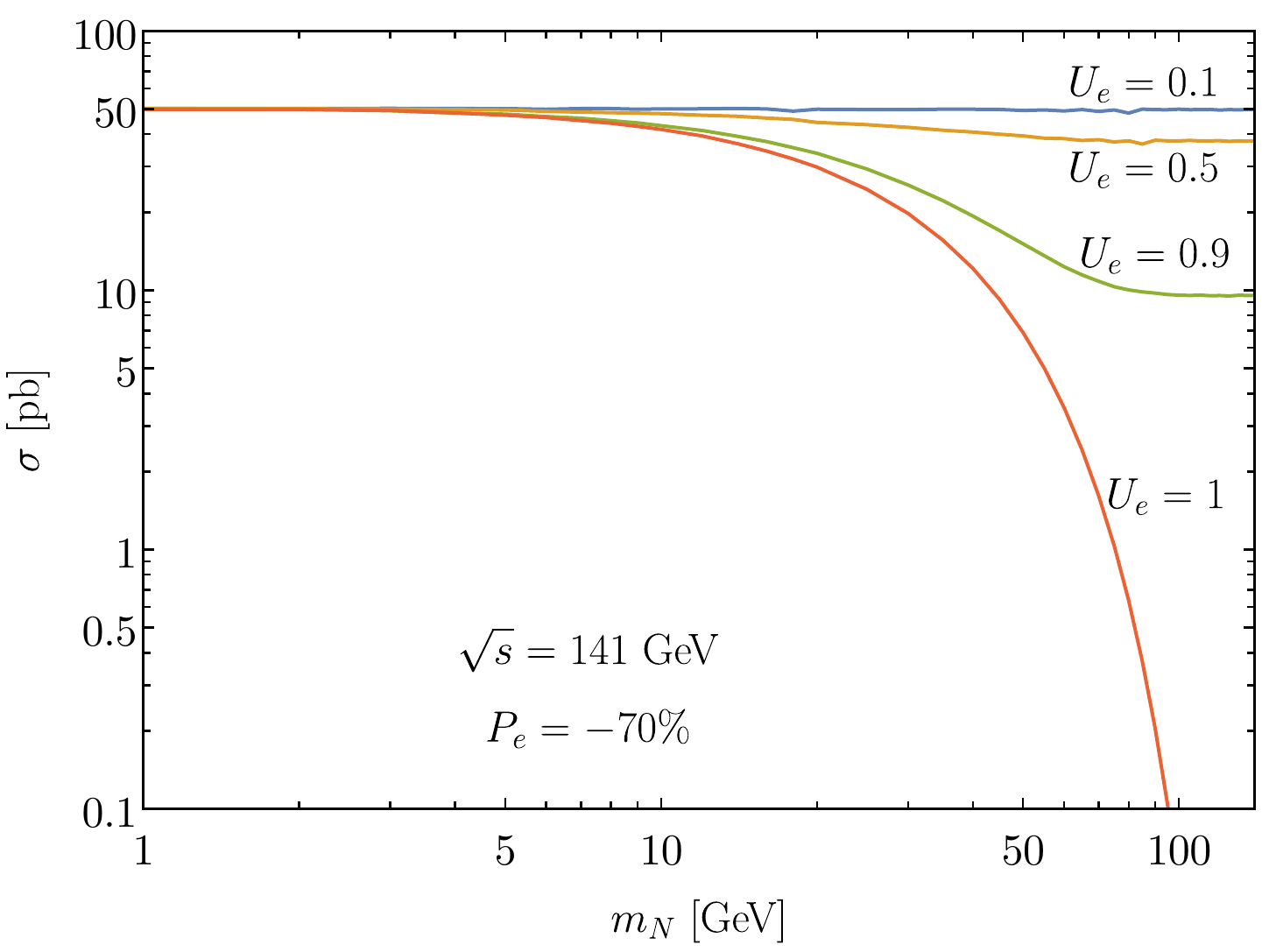}
		\caption{The cross section of mono-jet production at the EIC with collision energy $\sqrt{s}=141$ GeV and electron beam polarization $P_e=-70\%$.}
		\label{fig:xsecmonojet}
	\end{figure}

	We now consider a scenario in which the HNL is undetected or
	decays to (quasi-)stable neutral particles, \emph{e.g.}, dark matter.
	In this case, HNL production will lead to the mono-jet signature of $ep\to j+\slashed{E}_T$. 
	The main background arises from the production of SM neutrinos, $ep\to j+\nu_e$.
	In Fig.~\ref{fig:dist_monojet}, we show the jet rapidity $\eta_j$ and transverse momentum $p_T^j$ distributions. 
	Unfortunately, we see no clear distinctions between the HNL production (signal) compared with the SM $\nu_e$ one (background) in this case, except a smaller cross section due to the threshold suppression for massive HNL. The only chance to infer the presence of the HNL lies in counting the total number of events provided the SM mono-jet rate can be precisely predicted.
	The total cross section for mono-jet events can be written as
	\begin{equation}
		\begin{aligned}\label{eq:monojet}
			\sigma(ep\to j+\slashed{E}_T)
			&=\sigma(ep\to j+\nu_e)+\sigma(ep\to j+N)\\
			&=\sigma_{\rm SM}(ep\to j+\nu_e)\left[(1-|U_e|^2)+|U_e|^2\Phi(m_N)\right].
		\end{aligned}
	\end{equation}
	Here the $ep\to j+N$ cross section is the same as that for $ep\to j+\nu_e$, except for the squared mixing angle factor $|U_e|^2$ and the phase space factor $\Phi$ which accounts for the effect of the nonzero HNL mass. 
	The factor $(1-|U_e|^2)$ reflects the reduction of the $W^+e\bar{\nu}_e$ coupling in the HNL model with respect to the SM.
	We show the cross section as a function of $m_N$ for a few representative values of $U_e$ in the left panel of Fig.~\ref{fig:xsecmonojet}. In contrast to Fig.~\ref{fig:xsec}, here we show the results for a $-70\%$ polarized electron beam, with the cross section enhanced by a factor of 1.7, as discussed earlier.
	In the massless limit, \emph{i.e.}, $m_N\to0$, the HNL phase space should be equal to that of the SM neutrino, so that $\Phi(m_N = 0)=1$. In another limit $|U_e|^2\to0$, we expect no HNL contribution. Both of these scenarios match the SM case. 
	Here we take an aggressive acceptance, similarly to Ref.~\cite{AbdulKhalek:2021gbh}, that with reconstruction from both charged and neutral particles, the transverse momentum of the jet can be extended to 0.25 GeV in the pseudorapidity region $|\eta|<3.5$.
	
\begin{figure}
		\centering
		\includegraphics[width=0.8\textwidth]{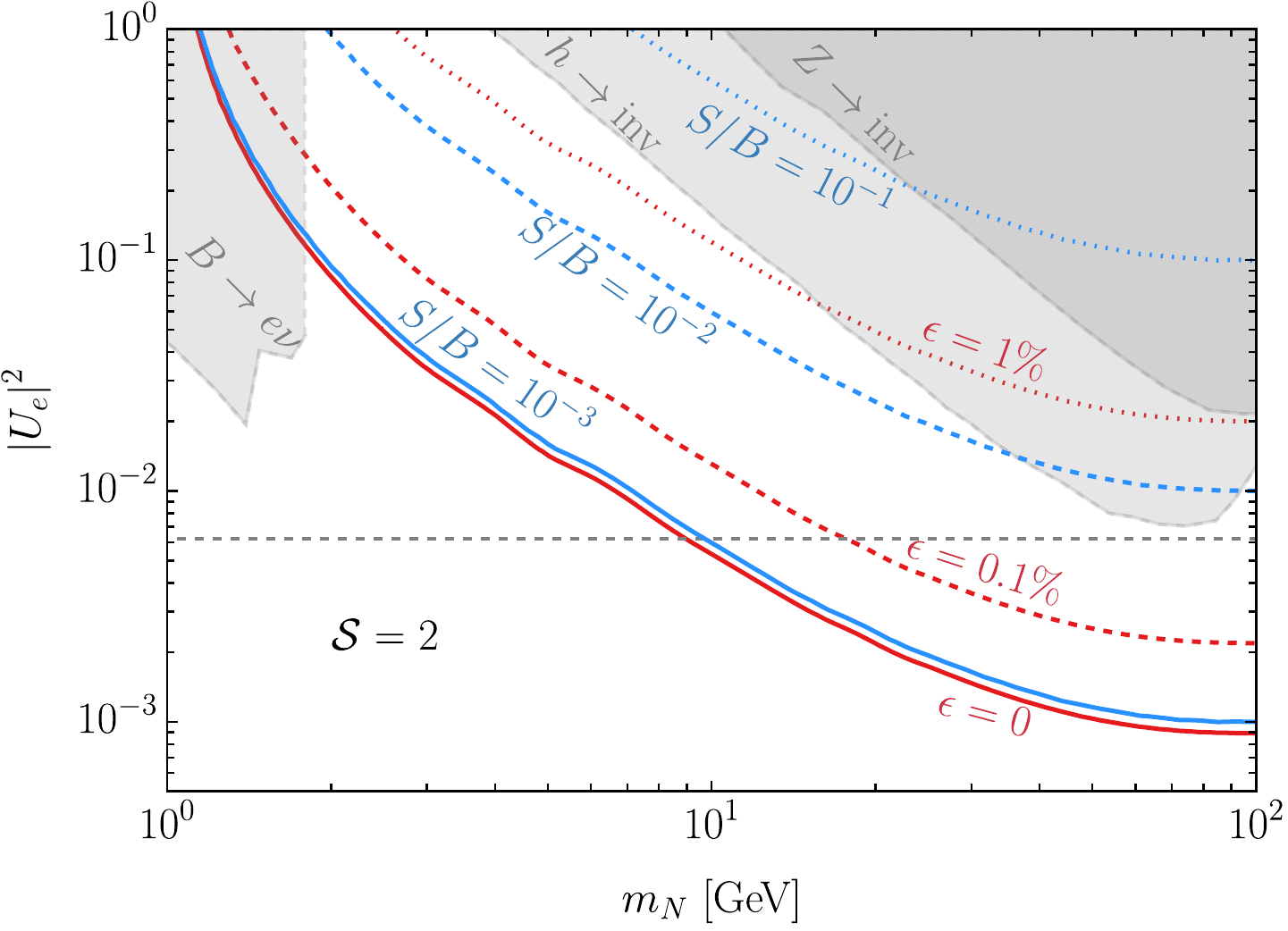}
		\caption{The sensitivity probe (red lines) of the EIC based on the mono-jet search, quantified with $\mathcal{S}=2$ in Eq.~(\ref{eq:sens}), with the relative systematic uncertainty as $\epsilon=0,~0.1\%$, and 1\%. The existing bounds come from invisible decays of $Z$ and Higgs bosons~\cite{CMS:2016dhk,Zyla:2020zbs}, peak searches in $B\rightarrow e\nu$ decays~\cite{Belle:2016nvh} (gray shaded) and indirect constraints from precision electroweak observables (dashed line)~\cite{MuLan:2012sih}. Also shown are contours of signal-to-background ratios $S/B=10^{-3},~10^{-2}$, and $10^{-1}$ (light blue lines).
		}
		\label{fig:monojet}
	\end{figure}

	We can define a statistical sensitivity to our HNL model as
	\begin{equation}\label{eq:sens}
		\mathcal{S}=\frac{S}{\sqrt{B+(\epsilon B)^2}},
	\end{equation}
	where
	\begin{equation}
		S=|N-N_{\rm SM}|, ~ B=N_{\rm SM},~
		N_{\rm (SM)} =\mathcal{L}\sigma_{\rm(SM)}.
	\end{equation}
	Here $\epsilon$ is the fractional error for systematic uncertainty with respect to the SM background events.
	The corresponding sensitivity in the two-dimensional plane $(m_N,|U_e|^2)$ is shown in the right panel of Fig.~\ref{fig:monojet}. Here we plot contours corresponding to $\mathcal{S}=2$, with assumed relative systematic uncertainties of 0, 0.1\%, and 1\%. We see that the sensitivity displays a strong dependence on the relative systematic uncertainty. For this reason, we also show the signal-to-background ratio $S/B$ of $10^{-3}$, $10^{-2}$, and $10^{-1}$ as light blue lines in Fig.~\ref{fig:monojet} as well.
	Based on the cross sections shown in Fig.~\ref{fig:xsec}, the SM background event is about $N_{\rm SM}\sim3\cdot10^{6}$. Due to this large background event number, the $\mathcal{S}=2$ can probe to $|U_e|^2\sim10^{-3}$ level, if we assume no systematics. With 0.1\% (1\%) systematic assumption, the sensitive region is narrowed down to $|U_e|^2\sim2\cdot10^{-3}$ ($2\cdot10^{-2}$) in the large $m_N$ region. When $m_N\to0$, the phase space factor $\Phi(m_N)\to1$ in Eq.~(\ref{eq:monojet}), and as seen in Fig.~\ref{fig:monojet} (left), the mono-jet cross section in HNL approaches to the SM one. Thus there is a gradual loss in sensitivity to $|U_e|^2$ for low values of $m_N$. 
	
	In Fig.~\ref{fig:monojet}, we compare our EIC projections in the mono-jet channel with other probes of invisible HNL decays, taking the existing bounds from Ref.~\cite{Batell:2017cmf}. Existing constraints on invisible $Z$ and Higgs decays~\cite{Zyla:2020zbs,CMS:2016dhk} cover part of the parameter space for heavier HNLs, while a peak search in the decay $B\rightarrow e \nu$ provides relevant constraints for GeV-scale HNLs~\cite{Belle:2016nvh}. There are also relatively strong, albeit indirect, constraints from precision electroweak tests over the full mass range, such as the MuLan bound~\cite{MuLan:2012sih} shown as the dashed line in Fig.~\ref{fig:monojet}.
	The ability of the EIC to compete with these existing constraints will depend to a large extent on how well systematic uncertainties can be brought under control, as is clearly seen in Fig.~\ref{fig:monojet}.
	
	\section{Discussion and Outlook}
	\label{sec:Disc}
	
	In this paper, we examined the feasibility of the EIC to search for new heavy neutral leptons produced in electron-proton collisions through charged current interactions as a consequence of their mixing with light SM neutrinos. 
	HNLs are well motivated due to their connections with neutrino mass generation and lepton number violation, as well as potentially offering a connection to a dark sector. 
	We studied several possible HNL signatures at the EIC, including prompt decays to visible final states, which are relevant for heavy HNLs with large mixing angles; displaced/long-lived particle signatures, which are predicted for light HNLs with small mixing angles; and purely invisible HNLs, which may occur if the HNL decays to invisible dark particles. These complementary signatures probe different HNL models/scenarios and regions of the HNL mass-mixing angle parameter space. 
	
\begin{figure}[tbh]
		\begin{center}
			\includegraphics[width=0.8\textwidth]{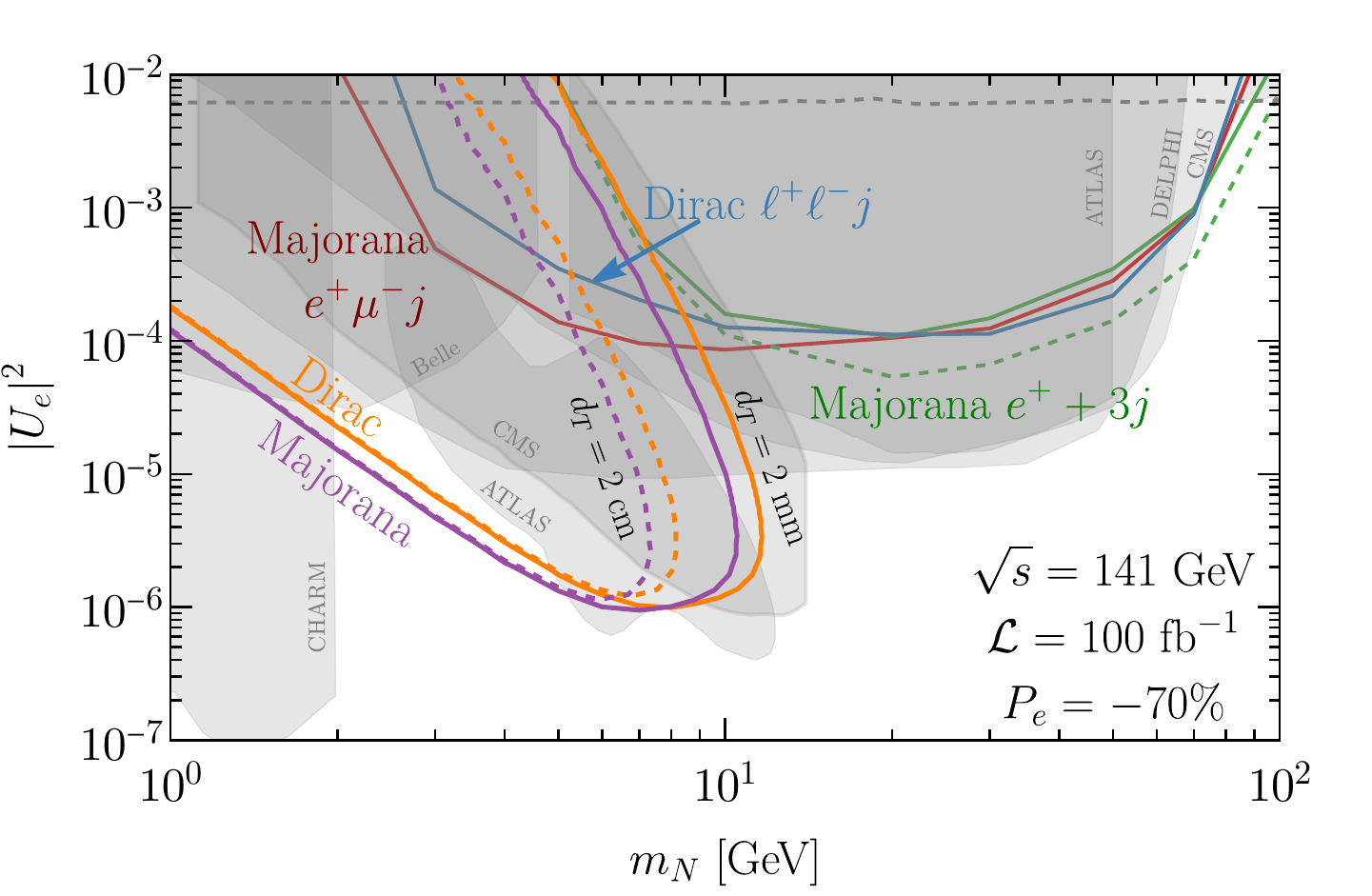}
		\end{center}
		\caption{The combined EIC sensitivity to HNL, compared with the existing bounds~\cite{CHARM:1985nku,DELPHI:1996qcc,CMS:2018iaf,CMS:2022fut,ATLAS:2019kpx,ATLAS:2022atq,Belle:2013ytx,MuLan:2012sih}.
			For details we refer the reader to Figs.~\ref{fig:prompt} and \ref{fig:DVS}.
		}
		\label{fig:sum}
	\end{figure}

	Our projections are derived using a detailed simulation of the production and decays of HNLs at the EIC that account for detector acceptance and resolutions. Suitable topological and kinematic cuts are applied to efficiently separate the HNL signal from the SM backgrounds. For prompt HNL decays, we analyzed both lepton-number-violating and conserving final states containing leptons $e,\mu$ in great detail. We found that 
	with the EIC design energy and integrated luminosity, one is able to probe the mass range of $1-100$ GeV and mixing angles of the order $10^{-4}-10^{-3}$. 
	Our results for these prompt-decay channels are summarized in Fig.~\ref{fig:prompt}.
	For a long-lived $N$ with a smaller mixing and lighter mass, we considered the distinctive signal of a displaced lepton with a large impact parameter, finding that the EIC can probe new territory in the few-GeV mass range for mixing angles of the order $10^{-6}-10^{-4}$. Our results for these displaced vertex channels at the $95\%$ C.L.~sensitivity are summarized in Fig.~\ref{fig:DVS}.
	The combined EIC sensitivity to HNL, compared with the existing bounds, is presented in a summary plot in Fig.~\ref{fig:sum}.
	For the invisible channel, where the HNL is undetected or decaying to the dark sector particle, one can potentially probe heavy HNLs for mixing angles in the window $10^{-3}-10^{-2}$, provided SM background systematics can be controlled. We summarize our results of $2\sigma$ sensitivity for the invisible decay via the monojet channel in Fig.~\ref{fig:monojet}.
	
	We now comment on several future avenues for investigation which may be fruitful. 
	With the development of effective $\tau$ tagging at the EIC, HNL decays to $\tau$ final states can also conceivably be exploited. 
	The displaced HNL signatures suggested in this work motivate careful consideration of the detector capabilities (\emph{e.g.},  tracking, angular coverage, and event timing, etc.) needed to exploit signatures of long-lived particles. Looking towards the future, there has been some discussion of a muon-ion collider at BNL following the EIC; see \emph{e.g.} Ref.~\cite{Acosta:2021qpx}. This higher energy machine would also allow for interesting probes of BSM physics, including HNLs with primarily muon-flavor mixing, and it would be worth exploring this in detail. 
	
	We have shown that the EIC can provide interesting probes of HNLs that are complementary to other experiments, such as neutrino-less double-$\beta$ decay, meson decays, and HNL production at fixed-target experiments and colliders.
	The strategies proposed here, with suitable adaptations, may be useful in other new physics searches, and we look forward to continued exploration of the potential of the EIC  to search for BSM physics. 
	
	\acknowledgments
	
	We thank Yulia Furletova for a helpful discussion about the EIC design during the early stage of this work. The work of BB, TH, and KX is supported by the U.S. Department of Energy under grant No. DE-SC0007914 and by PITT PACC. KX is also supported by the U.S. National Science Foundation under Grants No. PHY-1820760. TG would like to acknowledge the support from the Department of Atomic Energy, Government of India for the Regional Centre for Accelerator-based Particle Physics (RECAPP) at Harish Chandra Research Institute.

	\bibliographystyle{utphys}
	\bibliography{EIC-HNL.bib}

\providecommand{\href}[2]{#2}\begingroup\raggedright\begin{thebibliography}{10}

\bibitem{Zyla:2020zbs}
{\bfseries Particle Data Group} Collaboration, P.~Zyla {\em et~al.}, ``{Review
  of Particle Physics},'' \href{http://dx.doi.org/10.1093/ptep/ptaa104}{{\em
  PTEP} {\bfseries 2020} no.~8, (2020) 083C01}.

\bibitem{Accardi:2012qut}
A.~Accardi {\em et~al.}, ``{Electron Ion Collider: The Next QCD Frontier}:
  {Understanding the glue that binds us all},''
  \href{http://dx.doi.org/10.1140/epja/i2016-16268-9}{{\em Eur. Phys. J. A}
  {\bfseries 52} no.~9, (2016) 268},
  \href{http://arxiv.org/abs/1212.1701}{{\ttfamily arXiv:1212.1701 [nucl-ex]}}.

\bibitem{AbdulKhalek:2021gbh}
R.~Abdul~Khalek {\em et~al.}, ``{Science Requirements and Detector Concepts for
  the Electron-Ion Collider: EIC Yellow Report},''
  \href{http://arxiv.org/abs/2103.05419}{{\ttfamily arXiv:2103.05419
  [physics.ins-det]}}.

\bibitem{EIC-detector-handbook}
E.~A. et~al., ``{Electron-Ion Collider Detector Requirements and R$\&$D
  Handbook}.'' Version 1.2, February 25, 2020.

\bibitem{Kumar:2016mfi}
K.~Kumar, A.~Deshpande, J.~Huang, S.~Riordan, and Y.~Zhao, ``{Electroweak and
  BSM Physics at the EIC},''
  \href{http://dx.doi.org/10.1051/epjconf/201611203004}{{\em EPJ Web Conf.}
  {\bfseries 112} (2016) 03004}.

\bibitem{Boughezal:2022pmb}
R.~Boughezal, A.~Emmert, T.~Kutz, S.~Mantry, M.~Nycz, F.~Petriello,
  K.~\c{S}im\c{s}ek, D.~Wiegand, and X.~Zheng, ``{Neutral-current electroweak
  physics and SMEFT studies at the EIC},''
  \href{http://dx.doi.org/10.1103/PhysRevD.106.016006}{{\em Phys. Rev. D}
  {\bfseries 106} no.~1, (2022) 016006},
  \href{http://arxiv.org/abs/2204.07557}{{\ttfamily arXiv:2204.07557
  [hep-ph]}}.

\bibitem{Kumar:2013yoa}
K.~Kumar, S.~Mantry, W.~Marciano, and P.~Souder, ``{Low Energy Measurements of
  the Weak Mixing Angle},''
  \href{http://dx.doi.org/10.1146/annurev-nucl-102212-170556}{{\em Ann. Rev.
  Nucl. Part. Sci.} {\bfseries 63} (2013) 237--267},
  \href{http://arxiv.org/abs/1302.6263}{{\ttfamily arXiv:1302.6263 [hep-ex]}}.

\bibitem{Zhao:2016rfu}
Y.~Zhao, A.~Deshpande, J.~Huang, K.~Kumar, and S.~Riordan, ``{Neutral-Current
  Weak Interactions at an EIC},''
  \href{http://dx.doi.org/10.1140/epja/i2017-12245-2}{{\em Eur. Phys. J. A}
  {\bfseries 53} no.~3, (2017) 55},
  \href{http://arxiv.org/abs/1612.06927}{{\ttfamily arXiv:1612.06927
  [nucl-ex]}}.

\bibitem{RamseyMusolf:1999qk}
M.~Ramsey-Musolf, ``{Low-energy parity violation and new physics},''
  \href{http://dx.doi.org/10.1103/PhysRevC.60.015501}{{\em Phys. Rev. C}
  {\bfseries 60} (1999) 015501},
  \href{http://arxiv.org/abs/hep-ph/9903264}{{\ttfamily arXiv:hep-ph/9903264}}.

\bibitem{Davoudiasl:2012ag}
H.~Davoudiasl, H.-S. Lee, and W.~J. Marciano, ``{'Dark' Z implications for
  Parity Violation, Rare Meson Decays, and Higgs Physics},''
  \href{http://dx.doi.org/10.1103/PhysRevD.85.115019}{{\em Phys. Rev. D}
  {\bfseries 85} (2012) 115019},
  \href{http://arxiv.org/abs/1203.2947}{{\ttfamily arXiv:1203.2947 [hep-ph]}}.

\bibitem{Erler:2014fqa}
J.~Erler, C.~J. Horowitz, S.~Mantry, and P.~A. Souder, ``{Weak Polarized
  Electron Scattering},''
  \href{http://dx.doi.org/10.1146/annurev-nucl-102313-025520}{{\em Ann. Rev.
  Nucl. Part. Sci.} {\bfseries 64} (2014) 269--298},
  \href{http://arxiv.org/abs/1401.6199}{{\ttfamily arXiv:1401.6199 [hep-ph]}}.

\bibitem{Yan:2022npz}
B.~Yan, ``{Probing the dark photon via polarized DIS scattering at the HERA and
  EIC},'' \href{http://arxiv.org/abs/2203.01510}{{\ttfamily arXiv:2203.01510
  [hep-ph]}}.

\bibitem{Gonderinger:2010yn}
M.~Gonderinger and M.~J. Ramsey-Musolf, ``{Electron-to-Tau Lepton Flavor
  Violation at the Electron-Ion Collider},''
  \href{http://dx.doi.org/10.1007/JHEP11(2010)045}{{\em JHEP} {\bfseries 11}
  (2010) 045}, \href{http://arxiv.org/abs/1006.5063}{{\ttfamily arXiv:1006.5063
  [hep-ph]}}. [Erratum: JHEP 05, 047 (2012)].

\bibitem{Adloff:1999tp}
{\bfseries H1} Collaboration, C.~Adloff {\em et~al.}, ``{A Search for
  leptoquark bosons and lepton flavor violation in $e^{+} p$ collisions at
  HERA},'' \href{http://dx.doi.org/10.1007/s100520050646}{{\em Eur. Phys. J. C}
  {\bfseries 11} (1999) 447--471},
  \href{http://arxiv.org/abs/hep-ex/9907002}{{\ttfamily arXiv:hep-ex/9907002}}.
  [Erratum: Eur.Phys.J.C 14, 553--554 (2000)].

\bibitem{Chekanov:2005au}
{\bfseries ZEUS} Collaboration, S.~Chekanov {\em et~al.}, ``{Search for
  lepton-flavor violation at HERA},''
  \href{http://dx.doi.org/10.1140/epjc/s2005-02399-1}{{\em Eur. Phys. J. C}
  {\bfseries 44} (2005) 463--479},
  \href{http://arxiv.org/abs/hep-ex/0501070}{{\ttfamily arXiv:hep-ex/0501070}}.

\bibitem{Davoudiasl:2021mjy}
H.~Davoudiasl, R.~Marcarelli, and E.~T. Neil, ``{Lepton-Flavor-Violating ALPs
  at the Electron-Ion Collider: A Golden Opportunity},''
  \href{http://arxiv.org/abs/2112.04513}{{\ttfamily arXiv:2112.04513
  [hep-ph]}}.

\bibitem{Liu:2021lan}
Y.~Liu and B.~Yan, ``{Searching for the axion-like particle at the EIC},''
  \href{http://arxiv.org/abs/2112.02477}{{\ttfamily arXiv:2112.02477
  [hep-ph]}}.

\bibitem{Abdullahi:2022jlv}
A.~M. Abdullahi {\em et~al.}, ``{The Present and Future Status of Heavy Neutral
  Leptons},'' in {\em {2022 Snowmass Summer Study}}.
\newblock 3, 2022.
\newblock \href{http://arxiv.org/abs/2203.08039}{{\ttfamily arXiv:2203.08039
  [hep-ph]}}.

\bibitem{Minkowski:1977sc}
P.~Minkowski, ``{$\mu \to e\gamma$ at a Rate of One Out of $10^{9}$ Muon
  Decays?},''
\href{http://dx.doi.org/10.1016/0370-2693(77)90435-X}{{\em Phys. Lett.}
  {\bfseries 67B} (1977) 421--428}.

\bibitem{Yanagida:1979as}
T.~Yanagida, ``{HORIZONTAL SYMMETRY AND MASSES OF NEUTRINOS},''
{\em Conf. Proc.} {\bfseries C7902131} (1979) 95--99.

\bibitem{GellMann:1980vs}
M.~Gell-Mann, P.~Ramond, and R.~Slansky, ``{Complex Spinors and Unified
  Theories},'' {\em Conf. Proc.} {\bfseries C790927} (1979) 315--321,
\href{http://arxiv.org/abs/1306.4669}{{\ttfamily arXiv:1306.4669 [hep-th]}}.

\bibitem{Glashow:1979nm}
S.~L. Glashow, ``{The Future of Elementary Particle Physics},''
\href{http://dx.doi.org/10.1007/978-1-4684-7197-7_15}{{\em NATO Sci. Ser. B}
  {\bfseries 61} (1980) 687}.

\bibitem{Mohapatra:1979ia}
R.~N. Mohapatra and G.~Senjanovic, ``{Neutrino Mass and Spontaneous Parity
  Violation},''
\href{http://dx.doi.org/10.1103/PhysRevLett.44.912}{{\em Phys. Rev. Lett.}
  {\bfseries 44} (1980) 912}.

\bibitem{Schechter:1980gr}
J.~Schechter and J.~W.~F. Valle, ``{Neutrino Masses in SU(2) x U(1)
  Theories},''
\href{http://dx.doi.org/10.1103/PhysRevD.22.2227}{{\em Phys. Rev.} {\bfseries
  D22} (1980) 2227}.

\bibitem{Mohapatra:1986aw}
R.~Mohapatra, ``{Mechanism for Understanding Small Neutrino Mass in Superstring
  Theories},'' \href{http://dx.doi.org/10.1103/PhysRevLett.56.561}{{\em Phys.
  Rev. Lett.} {\bfseries 56} (1986) 561--563}.

\bibitem{Mohapatra:1986bd}
R.~Mohapatra and J.~Valle, ``{Neutrino Mass and Baryon Number Nonconservation
  in Superstring Models},''
  \href{http://dx.doi.org/10.1103/PhysRevD.34.1642}{{\em Phys. Rev. D}
  {\bfseries 34} (1986) 1642}.

\bibitem{Bernabeu:1987gr}
J.~Bernabeu, A.~Santamaria, J.~Vidal, A.~Mendez, and J.~Valle, ``{Lepton Flavor
  Nonconservation at High-Energies in a Superstring Inspired Standard Model},''
  \href{http://dx.doi.org/10.1016/0370-2693(87)91100-2}{{\em Phys. Lett. B}
  {\bfseries 187} (1987) 303--308}.

\bibitem{Malinsky:2005bi}
M.~Malinsky, J.~Romao, and J.~Valle, ``{Novel supersymmetric SO(10) seesaw
  mechanism},'' \href{http://dx.doi.org/10.1103/PhysRevLett.95.161801}{{\em
  Phys. Rev. Lett.} {\bfseries 95} (2005) 161801},
  \href{http://arxiv.org/abs/hep-ph/0506296}{{\ttfamily arXiv:hep-ph/0506296}}.

\bibitem{Kim:2020vjv}
Y.-H. Kim, ``{Neutrinoless double beta decay experiment},''
  \href{http://arxiv.org/abs/2004.02510}{{\ttfamily arXiv:2004.02510
  [hep-ex]}}.

\bibitem{Atre:2009rg}
A.~Atre, T.~Han, S.~Pascoli, and B.~Zhang, ``{The Search for Heavy Majorana
  Neutrinos},'' \href{http://dx.doi.org/10.1088/1126-6708/2009/05/030}{{\em
  JHEP} {\bfseries 05} (2009) 030},
  \href{http://arxiv.org/abs/0901.3589}{{\ttfamily arXiv:0901.3589 [hep-ph]}}.

\bibitem{Cai:2017mow}
Y.~Cai, T.~Han, T.~Li, and R.~Ruiz, ``{Lepton Number Violation: Seesaw Models
  and Their Collider Tests},''
  \href{http://dx.doi.org/10.3389/fphy.2018.00040}{{\em Front. in Phys.}
  {\bfseries 6} (2018) 40}, \href{http://arxiv.org/abs/1711.02180}{{\ttfamily
  arXiv:1711.02180 [hep-ph]}}.

\bibitem{Valle:1983dk}
J.~Valle and M.~Singer, ``{Lepton Number Violation With Quasi Dirac
  Neutrinos},'' \href{http://dx.doi.org/10.1103/PhysRevD.28.540}{{\em Phys.
  Rev. D} {\bfseries 28} (1983) 540}.

\bibitem{Asaka:2005pn}
T.~Asaka and M.~Shaposhnikov, ``{The $\nu$MSM, dark matter and baryon asymmetry
  of the universe},''
  \href{http://dx.doi.org/10.1016/j.physletb.2005.06.020}{{\em Phys. Lett. B}
  {\bfseries 620} (2005) 17--26},
  \href{http://arxiv.org/abs/hep-ph/0505013}{{\ttfamily arXiv:hep-ph/0505013}}.

\bibitem{Akhmedov:1998qx}
E.~K. Akhmedov, V.~Rubakov, and A.~Smirnov, ``{Baryogenesis via neutrino
  oscillations},'' \href{http://dx.doi.org/10.1103/PhysRevLett.81.1359}{{\em
  Phys. Rev. Lett.} {\bfseries 81} (1998) 1359--1362},
  \href{http://arxiv.org/abs/hep-ph/9803255}{{\ttfamily arXiv:hep-ph/9803255}}.

\bibitem{Pospelov:2007mp}
M.~Pospelov, A.~Ritz, and M.~B. Voloshin, ``{Secluded WIMP Dark Matter},''
  \href{http://dx.doi.org/10.1016/j.physletb.2008.02.052}{{\em Phys. Lett. B}
  {\bfseries 662} (2008) 53--61},
  \href{http://arxiv.org/abs/0711.4866}{{\ttfamily arXiv:0711.4866 [hep-ph]}}.

\bibitem{Bertoni:2014mva}
B.~Bertoni, S.~Ipek, D.~McKeen, and A.~E. Nelson, ``{Constraints and
  consequences of reducing small scale structure via large dark matter-neutrino
  interactions},'' \href{http://dx.doi.org/10.1007/JHEP04(2015)170}{{\em JHEP}
  {\bfseries 04} (2015) 170}, \href{http://arxiv.org/abs/1412.3113}{{\ttfamily
  arXiv:1412.3113 [hep-ph]}}.

\bibitem{Gonzalez-Macias:2016vxy}
V.~Gonz\'alez-Mac\'\i{}as, J.~I. Illana, and J.~Wudka, ``{A realistic model for
  Dark Matter interactions in the neutrino portal paradigm},''
  \href{http://dx.doi.org/10.1007/JHEP05(2016)171}{{\em JHEP} {\bfseries 05}
  (2016) 171}, \href{http://arxiv.org/abs/1601.05051}{{\ttfamily
  arXiv:1601.05051 [hep-ph]}}.

\bibitem{Escudero:2016ksa}
M.~Escudero, N.~Rius, and V.~Sanz, ``{Sterile Neutrino portal to Dark Matter
  II: Exact Dark symmetry},''
  \href{http://dx.doi.org/10.1140/epjc/s10052-017-4963-x}{{\em Eur. Phys. J. C}
  {\bfseries 77} no.~6, (2017) 397},
  \href{http://arxiv.org/abs/1607.02373}{{\ttfamily arXiv:1607.02373
  [hep-ph]}}.

\bibitem{Batell:2017rol}
B.~Batell, T.~Han, and B.~Shams Es~Haghi, ``{Indirect Detection of Neutrino
  Portal Dark Matter},''
  \href{http://dx.doi.org/10.1103/PhysRevD.97.095020}{{\em Phys. Rev. D}
  {\bfseries 97} no.~9, (2018) 095020},
  \href{http://arxiv.org/abs/1704.08708}{{\ttfamily arXiv:1704.08708
  [hep-ph]}}.

\bibitem{Batell:2017cmf}
B.~Batell, T.~Han, D.~McKeen, and B.~Shams Es~Haghi, ``{Thermal Dark Matter
  Through the Dirac Neutrino Portal},''
  \href{http://dx.doi.org/10.1103/PhysRevD.97.075016}{{\em Phys. Rev. D}
  {\bfseries 97} no.~7, (2018) 075016},
  \href{http://arxiv.org/abs/1709.07001}{{\ttfamily arXiv:1709.07001
  [hep-ph]}}.

\bibitem{Schmaltz:2017oov}
M.~Schmaltz and N.~Weiner, ``{A Portalino to the Dark Sector},''
  \href{http://dx.doi.org/10.1007/JHEP02(2019)105}{{\em JHEP} {\bfseries 02}
  (2019) 105}, \href{http://arxiv.org/abs/1709.09164}{{\ttfamily
  arXiv:1709.09164 [hep-ph]}}.

\bibitem{Hou:2019efy}
T.-J. Hou {\em et~al.}, ``{New CTEQ global analysis of quantum chromodynamics
  with high-precision data from the LHC},''
  \href{http://dx.doi.org/10.1103/PhysRevD.103.014013}{{\em Phys. Rev. D}
  {\bfseries 103} no.~1, (2021) 014013},
  \href{http://arxiv.org/abs/1912.10053}{{\ttfamily arXiv:1912.10053
  [hep-ph]}}.

\bibitem{Buchmuller:1990vh}
W.~Buchmuller and C.~Greub, ``{Electroproduction of Majorana neutrinos},''
  \href{http://dx.doi.org/10.1016/0370-2693(91)91792-T}{{\em Phys. Lett. B}
  {\bfseries 256} (1991) 465--470}.

\bibitem{Buchmuller:1991tu}
W.~Buchmuller and C.~Greub, ``{Heavy Majorana neutrinos in electron - positron
  and electron - proton collisions},''
  \href{http://dx.doi.org/10.1016/0550-3213(91)80024-G}{{\em Nucl. Phys. B}
  {\bfseries 363} (1991) 345--368}.

\bibitem{Ingelman:1993ve}
G.~Ingelman and J.~Rathsman, ``{Heavy Majorana neutrinos at e p colliders},''
  \href{http://dx.doi.org/10.1007/BF01474620}{{\em Z. Phys. C} {\bfseries 60}
  (1993) 243--254}.

\bibitem{Das:2018usr}
A.~Das, S.~Jana, S.~Mandal, and S.~Nandi, ``{Probing right handed neutrinos at
  the LHeC and lepton colliders using fat jet signatures},''
  \href{http://dx.doi.org/10.1103/PhysRevD.99.055030}{{\em Phys. Rev. D}
  {\bfseries 99} no.~5, (2019) 055030},
  \href{http://arxiv.org/abs/1811.04291}{{\ttfamily arXiv:1811.04291
  [hep-ph]}}.

\bibitem{Li:2018wut}
S.-Y. Li, Z.-G. Si, and X.-H. Yang, ``{Heavy Majorana Neutrino Production at
  Future $ep$ Colliders},''
  \href{http://dx.doi.org/10.1016/j.physletb.2019.06.001}{{\em Phys. Lett. B}
  {\bfseries 795} (2019) 49--55},
  \href{http://arxiv.org/abs/1811.10313}{{\ttfamily arXiv:1811.10313
  [hep-ph]}}.

\bibitem{Gu:2022muc}
H.~Gu and K.~Wang, ``{Search for heavy Majorana neutrinos at electron-proton
  colliders},'' \href{http://dx.doi.org/10.1103/PhysRevD.106.015006}{{\em Phys.
  Rev. D} {\bfseries 106} no.~1, (2022) 015006},
  \href{http://arxiv.org/abs/2201.12997}{{\ttfamily arXiv:2201.12997
  [hep-ph]}}.

\bibitem{Giffin:2022rei}
P.~Giffin, S.~Gori, Y.-D. Tsai, and D.~Tuckler, ``{Heavy Neutral Leptons at
  Beam Dump Experiments of Future Lepton Colliders},''
  \href{http://arxiv.org/abs/2206.13745}{{\ttfamily arXiv:2206.13745
  [hep-ph]}}.

\bibitem{Sherpa:2019gpd}
{\bfseries Sherpa} Collaboration, E.~Bothmann {\em et~al.}, ``{Event Generation
  with Sherpa 2.2}'' \href{http://dx.doi.org/10.21468/SciPostPhys.7.3.034}{{\em
  SciPost Phys.} {\bfseries 7} no.~3, (2019) 034},
  \href{http://arxiv.org/abs/1905.09127}{{\ttfamily arXiv:1905.09127
  [hep-ph]}}.

\bibitem{Nocera:2014gqa}
{\bfseries NNPDF} Collaboration, E.~R. Nocera, R.~D. Ball, S.~Forte,
  G.~Ridolfi, and J.~Rojo, ``{A first unbiased global determination of
  polarized PDFs and their uncertainties},''
  \href{http://dx.doi.org/10.1016/j.nuclphysb.2014.08.008}{{\em Nucl. Phys. B}
  {\bfseries 887} (2014) 276--308},
  \href{http://arxiv.org/abs/1406.5539}{{\ttfamily arXiv:1406.5539 [hep-ph]}}.

\bibitem{Gorbunov:2007ak}
D.~Gorbunov and M.~Shaposhnikov, ``{How to find neutral leptons of the
  $\nu$MSM?},'' \href{http://dx.doi.org/10.1088/1126-6708/2007/10/015}{{\em
  JHEP} {\bfseries 10} (2007) 015},
  \href{http://arxiv.org/abs/0705.1729}{{\ttfamily arXiv:0705.1729 [hep-ph]}}.
  [Erratum: JHEP 11, 101 (2013)].

\bibitem{Canetti:2012kh}
L.~Canetti, M.~Drewes, T.~Frossard, and M.~Shaposhnikov, ``{Dark Matter,
  Baryogenesis and Neutrino Oscillations from Right Handed Neutrinos},''
  \href{http://dx.doi.org/10.1103/PhysRevD.87.093006}{{\em Phys. Rev. D}
  {\bfseries 87} (2013) 093006},
  \href{http://arxiv.org/abs/1208.4607}{{\ttfamily arXiv:1208.4607 [hep-ph]}}.

\bibitem{Shuve:2016muy}
B.~Shuve and M.~E. Peskin, ``{Revision of the LHCb Limit on Majorana
  Neutrinos},'' \href{http://dx.doi.org/10.1103/PhysRevD.94.113007}{{\em Phys.
  Rev. D} {\bfseries 94} no.~11, (2016) 113007},
  \href{http://arxiv.org/abs/1607.04258}{{\ttfamily arXiv:1607.04258
  [hep-ph]}}.

\bibitem{Bondarenko:2018ptm}
K.~Bondarenko, A.~Boyarsky, D.~Gorbunov, and O.~Ruchayskiy, ``{Phenomenology of
  GeV-scale Heavy Neutral Leptons},''
  \href{http://dx.doi.org/10.1007/JHEP11(2018)032}{{\em JHEP} {\bfseries 11}
  (2018) 032}, \href{http://arxiv.org/abs/1805.08567}{{\ttfamily
  arXiv:1805.08567 [hep-ph]}}.

\bibitem{Arratia:2020azl}
M.~Arratia, Y.~Furletova, T.~Hobbs, F.~Olness, and S.~J. Sekula, ``{Charm jets
  as a probe for strangeness at the future Electron-Ion Collider},''
  \href{http://arxiv.org/abs/2006.12520}{{\ttfamily arXiv:2006.12520
  [hep-ph]}}.

\bibitem{Duarte:2014zea}
L.~Duarte, G.~A. Gonz\'alez-Sprinberg, and O.~A. Sampayo, ``{Majorana neutrinos
  production at LHeC in an effective approach},''
  \href{http://dx.doi.org/10.1103/PhysRevD.91.053007}{{\em Phys. Rev. D}
  {\bfseries 91} no.~5, (2015) 053007},
  \href{http://arxiv.org/abs/1412.1433}{{\ttfamily arXiv:1412.1433 [hep-ph]}}.

\bibitem{Antusch:2016ejd}
S.~Antusch, E.~Cazzato, and O.~Fischer, ``{Sterile neutrino searches at future
  $e^-e^+$, $pp$, and $e^-p$ colliders},''
  \href{http://dx.doi.org/10.1142/S0217751X17500786}{{\em Int. J. Mod. Phys. A}
  {\bfseries 32} no.~14, (2017) 1750078},
  \href{http://arxiv.org/abs/1612.02728}{{\ttfamily arXiv:1612.02728
  [hep-ph]}}.

\bibitem{Alva:2014gxa}
D.~Alva, T.~Han, and R.~Ruiz, ``{Heavy Majorana neutrinos from $W\gamma$ fusion
  at hadron colliders},'' \href{http://dx.doi.org/10.1007/JHEP02(2015)072}{{\em
  JHEP} {\bfseries 02} (2015) 072},
  \href{http://arxiv.org/abs/1411.7305}{{\ttfamily arXiv:1411.7305 [hep-ph]}}.

\bibitem{Degrande:2016aje}
C.~Degrande, O.~Mattelaer, R.~Ruiz, and J.~Turner, ``{Fully-Automated Precision
  Predictions for Heavy Neutrino Production Mechanisms at Hadron Colliders},''
  \href{http://dx.doi.org/10.1103/PhysRevD.94.053002}{{\em Phys. Rev. D}
  {\bfseries 94} no.~5, (2016) 053002},
  \href{http://arxiv.org/abs/1602.06957}{{\ttfamily arXiv:1602.06957
  [hep-ph]}}.

\bibitem{Pascoli:2018heg}
S.~Pascoli, R.~Ruiz, and C.~Weiland, ``{Heavy neutrinos with dynamic jet
  vetoes: multilepton searches at $ \sqrt{s}=14 $ , 27, and 100 TeV},''
  \href{http://dx.doi.org/10.1007/JHEP06(2019)049}{{\em JHEP} {\bfseries 06}
  (2019) 049}, \href{http://arxiv.org/abs/1812.08750}{{\ttfamily
  arXiv:1812.08750 [hep-ph]}}.

\bibitem{Alwall:2014hca}
J.~Alwall, R.~Frederix, S.~Frixione, V.~Hirschi, F.~Maltoni, O.~Mattelaer,
  H.~S. Shao, T.~Stelzer, P.~Torrielli, and M.~Zaro, ``{The automated
  computation of tree-level and next-to-leading order differential cross
  sections, and their matching to parton shower simulations},''
  \href{http://dx.doi.org/10.1007/JHEP07(2014)079}{{\em JHEP} {\bfseries 07}
  (2014) 079}, \href{http://arxiv.org/abs/1405.0301}{{\ttfamily arXiv:1405.0301
  [hep-ph]}}.

\bibitem{ATLAS:2019qmc}
{\bfseries ATLAS} Collaboration, G.~Aad {\em et~al.}, ``{Electron and photon
  performance measurements with the ATLAS detector using the
  2015\textendash{}2017 LHC proton-proton collision data},''
  \href{http://dx.doi.org/10.1088/1748-0221/14/12/P12006}{{\em JINST}
  {\bfseries 14} no.~12, (2019) P12006},
  \href{http://arxiv.org/abs/1908.00005}{{\ttfamily arXiv:1908.00005
  [hep-ex]}}.

\bibitem{Budnev:1975poe}
V.~M. Budnev, I.~F. Ginzburg, G.~V. Meledin, and V.~G. Serbo, ``{The Two photon
  particle production mechanism. Physical problems. Applications. Equivalent
  photon approximation},''
  \href{http://dx.doi.org/10.1016/0370-1573(75)90009-5}{{\em Phys. Rept.}
  {\bfseries 15} (1975) 181--281}.

\bibitem{CHARM:1985nku}
{\bfseries CHARM} Collaboration, F.~Bergsma {\em et~al.}, ``{A Search for
  Decays of Heavy Neutrinos in the Mass Range 0.5-{GeV} to 2.8-{GeV}},''
  \href{http://dx.doi.org/10.1016/0370-2693(86)91601-1}{{\em Phys. Lett. B}
  {\bfseries 166} (1986) 473--478}.

\bibitem{DELPHI:1996qcc}
{\bfseries DELPHI} Collaboration, P.~Abreu {\em et~al.}, ``{Search for neutral
  heavy leptons produced in Z decays},''
  \href{http://dx.doi.org/10.1007/s002880050370}{{\em Z. Phys. C} {\bfseries
  74} (1997) 57--71}. [Erratum: Z.Phys.C 75, 580 (1997)].

\bibitem{Belle:2013ytx}
{\bfseries Belle} Collaboration, D.~Liventsev {\em et~al.}, ``{Search for heavy
  neutrinos at Belle},''
  \href{http://dx.doi.org/10.1103/PhysRevD.87.071102}{{\em Phys. Rev. D}
  {\bfseries 87} no.~7, (2013) 071102},
  \href{http://arxiv.org/abs/1301.1105}{{\ttfamily arXiv:1301.1105 [hep-ex]}}.
  [Erratum: Phys.Rev.D 95, 099903 (2017)].

\bibitem{CMS:2018iaf}
{\bfseries CMS} Collaboration, A.~M. Sirunyan {\em et~al.}, ``{Search for heavy
  neutral leptons in events with three charged leptons in proton-proton
  collisions at $\sqrt{s} =$ 13 TeV},''
  \href{http://dx.doi.org/10.1103/PhysRevLett.120.221801}{{\em Phys. Rev.
  Lett.} {\bfseries 120} no.~22, (2018) 221801},
  \href{http://arxiv.org/abs/1802.02965}{{\ttfamily arXiv:1802.02965
  [hep-ex]}}.

\bibitem{CMS:2022fut}
{\bfseries CMS} Collaboration, A.~Tumasyan {\em et~al.}, ``{Search for
  long-lived heavy neutral leptons with displaced vertices in proton-proton
  collisions at $ \sqrt{\mathrm{s}} $ =13 TeV},''
  \href{http://dx.doi.org/10.1007/JHEP07(2022)081}{{\em JHEP} {\bfseries 07}
  (2022) 081}, \href{http://arxiv.org/abs/2201.05578}{{\ttfamily
  arXiv:2201.05578 [hep-ex]}}.

\bibitem{ATLAS:2019kpx}
{\bfseries ATLAS} Collaboration, G.~Aad {\em et~al.}, ``{Search for heavy
  neutral leptons in decays of $W$ bosons produced in 13 TeV $pp$ collisions
  using prompt and displaced signatures with the ATLAS detector},''
  \href{http://dx.doi.org/10.1007/JHEP10(2019)265}{{\em JHEP} {\bfseries 10}
  (2019) 265}, \href{http://arxiv.org/abs/1905.09787}{{\ttfamily
  arXiv:1905.09787 [hep-ex]}}.

\bibitem{ATLAS:2022atq}
{\bfseries ATLAS} Collaboration, ``{Search for heavy neutral leptons in decays
  of $W$ bosons using a dilepton displaced vertex in $\sqrt{s}=13$ TeV $pp$
  collisions with the ATLAS detector},''
  \href{http://arxiv.org/abs/2204.11988}{{\ttfamily arXiv:2204.11988
  [hep-ex]}}.

\bibitem{MuLan:2012sih}
{\bfseries MuLan} Collaboration, V.~Tishchenko {\em et~al.}, ``{Detailed Report
  of the MuLan Measurement of the Positive Muon Lifetime and Determination of
  the Fermi Constant},''
  \href{http://dx.doi.org/10.1103/PhysRevD.87.052003}{{\em Phys. Rev. D}
  {\bfseries 87} no.~5, (2013) 052003},
  \href{http://arxiv.org/abs/1211.0960}{{\ttfamily arXiv:1211.0960 [hep-ex]}}.

\bibitem{delAguila:2008pw}
F.~del Aguila, J.~de~Blas, and M.~Perez-Victoria, ``{Effects of new leptons in
  Electroweak Precision Data},''
  \href{http://dx.doi.org/10.1103/PhysRevD.78.013010}{{\em Phys. Rev. D}
  {\bfseries 78} (2008) 013010},
  \href{http://arxiv.org/abs/0803.4008}{{\ttfamily arXiv:0803.4008 [hep-ph]}}.

\bibitem{Akhmedov:2013hec}
E.~Akhmedov, A.~Kartavtsev, M.~Lindner, L.~Michaels, and J.~Smirnov,
  ``{Improving Electro-Weak Fits with TeV-scale Sterile Neutrinos},''
  \href{http://dx.doi.org/10.1007/JHEP05(2013)081}{{\em JHEP} {\bfseries 05}
  (2013) 081}, \href{http://arxiv.org/abs/1302.1872}{{\ttfamily arXiv:1302.1872
  [hep-ph]}}.

\bibitem{Basso:2013jka}
L.~Basso, O.~Fischer, and J.~J. van~der Bij, ``{Precision tests of unitarity in
  leptonic mixing},'' \href{http://dx.doi.org/10.1209/0295-5075/105/11001}{{\em
  EPL} {\bfseries 105} no.~1, (2014) 11001},
  \href{http://arxiv.org/abs/1310.2057}{{\ttfamily arXiv:1310.2057 [hep-ph]}}.

\bibitem{deBlas:2013gla}
J.~de~Blas, ``{Electroweak limits on physics beyond the Standard Model},''
  \href{http://dx.doi.org/10.1051/epjconf/20136019008}{{\em EPJ Web Conf.}
  {\bfseries 60} (2013) 19008},
  \href{http://arxiv.org/abs/1307.6173}{{\ttfamily arXiv:1307.6173 [hep-ph]}}.

\bibitem{Antusch:2015mia}
S.~Antusch and O.~Fischer, ``{Testing sterile neutrino extensions of the
  Standard Model at future lepton colliders},''
  \href{http://dx.doi.org/10.1007/JHEP05(2015)053}{{\em JHEP} {\bfseries 05}
  (2015) 053}, \href{http://arxiv.org/abs/1502.05915}{{\ttfamily
  arXiv:1502.05915 [hep-ph]}}.

\bibitem{Deppisch:2015qwa}
F.~F. Deppisch, P.~S. Bhupal~Dev, and A.~Pilaftsis, ``{Neutrinos and Collider
  Physics},'' \href{http://dx.doi.org/10.1088/1367-2630/17/7/075019}{{\em New
  J. Phys.} {\bfseries 17} no.~7, (2015) 075019},
  \href{http://arxiv.org/abs/1502.06541}{{\ttfamily arXiv:1502.06541
  [hep-ph]}}.

\bibitem{Drewes:2019fou}
M.~Drewes and J.~Hajer, ``{Heavy Neutrinos in displaced vertex searches at the
  LHC and HL-LHC},'' \href{http://dx.doi.org/10.1007/JHEP02(2020)070}{{\em
  JHEP} {\bfseries 02} (2020) 070},
  \href{http://arxiv.org/abs/1903.06100}{{\ttfamily arXiv:1903.06100
  [hep-ph]}}.

\bibitem{Liu:2019ayx}
J.~Liu, Z.~Liu, L.-T. Wang, and X.-P. Wang, ``{Seeking for sterile neutrinos
  with displaced leptons at the LHC},''
  \href{http://dx.doi.org/10.1007/JHEP07(2019)159}{{\em JHEP} {\bfseries 07}
  (2019) 159}, \href{http://arxiv.org/abs/1904.01020}{{\ttfamily
  arXiv:1904.01020 [hep-ph]}}.

\bibitem{CMS:2016dhk}
{\bfseries CMS} Collaboration, V.~Khachatryan {\em et~al.}, ``{Searches for
  invisible decays of the Higgs boson in pp collisions at $\sqrt{s}$ = 7, 8,
  and 13 TeV},'' \href{http://dx.doi.org/10.1007/JHEP02(2017)135}{{\em JHEP}
  {\bfseries 02} (2017) 135}, \href{http://arxiv.org/abs/1610.09218}{{\ttfamily
  arXiv:1610.09218 [hep-ex]}}.

\bibitem{Belle:2016nvh}
{\bfseries Belle} Collaboration, C.~S. Park {\em et~al.}, ``{Search for a
  massive invisible particle $X^0$ in $B^{+}\to e^{+}X^{0}$ and $B^{+}\to
  \mu^{+}X^{0}$ decays},''
  \href{http://dx.doi.org/10.1103/PhysRevD.94.012003}{{\em Phys. Rev. D}
  {\bfseries 94} no.~1, (2016) 012003},
  \href{http://arxiv.org/abs/1605.04430}{{\ttfamily arXiv:1605.04430
  [hep-ex]}}.

\bibitem{Acosta:2021qpx}
D.~Acosta and W.~Li, ``{A muon\textendash{}ion collider at BNL: The future QCD
  frontier and path to a new energy frontier of
  \ensuremath{\mu}+\ensuremath{\mu}\ensuremath{-} colliders},''
  \href{http://dx.doi.org/10.1016/j.nima.2022.166334}{{\em Nucl. Instrum. Meth.
  A} {\bfseries 1027} (2022) 166334},
  \href{http://arxiv.org/abs/2107.02073}{{\ttfamily arXiv:2107.02073
  [physics.acc-ph]}}.

\end{thebibliography}\endgroup

\end{document}